\documentclass[fleqn,10pt]{wlscirep}
\usepackage[utf8]{inputenc}
\usepackage[T1]{fontenc}
\usepackage[font=small,skip=0pt]{caption}
\usepackage{todonotes}
\usepackage{cite}
\usepackage{multirow}
\usepackage{amsmath,amssymb,amsfonts}
\usepackage{algorithmic}
\usepackage{graphicx}
\usepackage{dblfloatfix}
\usepackage{textcomp}
\usepackage{placeins}
\usepackage{xcolor}
\usepackage{rsfso}
\usepackage{rotating}

\graphicspath{{Figures/}}

\title{Multilevel Structural Evaluation of Signed Directed Social Networks based on Balance Theory}

\author[1,+,*]{Samin Aref}
\author[2,+,*]{Ly Dinh}
\author[2,+,*]{Rezvaneh Rezapour} 
\author[2]{Jana Diesner}
\affil[1]{Laboratory of Digital and Computational Demography, Max Planck Institute for Demographic Research, 18057, Rostock, Germany}
\affil[2]{School of Information Sciences, University of Illinois at Urbana-Champaign, Champaign, USA}

\affil[*]{Corresponding authors: aref@demogr.mpg.de, \{dinh4,rezapou2\}@illinois.edu}

\affil[+]{These authors contributed equally to this work.}

\begin{abstract}
	Balance theory explains the forces behind the structure of social systems, which are commonly modeled as static undirected signed networks. We expand this modeling approach to incorporate directionality of edges, and consider three levels of analysis: triads, subgroups, and the whole network. For triad-level balance, we operationalize a new measure by utilizing semicycles that satisfy the condition of transitivity. For subgroup-level balance, we propose measures of cohesiveness (intra-group solidarity) and divisiveness (inter-group antagonism) to capture balance within and among subgroups of the network using the most fitting partition of nodes into two groups. For network-level balance, we re-purpose the normalized line index to incorporate directionality, and provide the proportion of edges whose position suits balance. Through extensive computational analysis, we quantify and analyze patterns of social structure in triads, subgroups, and the whole network across a range of social settings from college students and Wikipedia editors to philosophers and Bitcoin traders. We then apply our multilevel framework of analysis to examine balance in temporal and multilayer networks, which demonstrates the generalizability of our approach to evaluating balance, and leads to new observations on balance with respect to time and layer dimensions. Our complementary findings on a variety of social networks highlight the need to evaluate balance at different levels. We propose a comprehensive yet parsimonious approach to address this need.
\end{abstract}
\begin{document}
	
	\flushbottom
	\maketitle

	\section*{Introduction}
	Social entities 
	can establish different types of relationships, such as personal or professional ones. These relationships might be reciprocal or not, and may have a dual opposite nature, such as affection versus dislike, friendship versus enmity, or trust versus distrust. Throughout time and across types of relationships, ties may change. Signed directed graphs can be used to model such complex relationships as network, where ties have two properties: a sign (positive or negative) and directionality (directed or not). Additionally, the type and time of relationships can be modeled with layering and temporal edge attributes. Analyzing the resulting data allows us to explore the structure and dynamics of relationships among individuals or collectives as social entities, elaborate observations based on social science theories, or put existing theories to empirical test.
	
	One existing theory is structural balance \cite{heider1946attitudes,cartwright_structural_1956}, which has been widely used to explain local-level social dynamics that may emerge within and between triads, potentially ripple through a network, and hence can lead to network-wide effects. With its root in social and cognitive psychology \cite{heider1946attitudes}, the theory explains how different configurations of positive and negative relationships between pairs of nodes may impact the amount of tension in a triad (three nodes with an edge between every pair). This tension would be rather absent if the triad has an even number of (0 or 2) negative edges \cite{cartwright_structural_1956}. Applying this premise to real-world situations, the following four adages scope out the balanced configurations at the triadic level based on edge signs: \textit{my friend’s friend is my friend (+++); my friend’s enemy is my enemy (+- -); my enemy’s friend is my enemy (-+-); and my enemy’s enemy is my friend (- -+).} The measurement of balance has later been expanded from the triads level (micro-level) to the subgroup level (meso-level) by partitioning nodes into two groups or ``plus-sets'' \cite{doreian_partitioning_1996,davis1967}, such that the number of positive edges between groups and negative edges within groups is minimized. The measurement of balance could also be expanded to the network level (macro-level) using the \textit{line index of balance} \cite{harary1959measurement,flament1970equilibre,zaslavsky_balanced_1987,aref2015measuring,aref2017balance}, which equals the minimum number of positive edges between groups and negative edges within groups across all possible ways to partition the nodes into two groups. Balance is widely assumed to emerge at the level of triads, such that changes within any triad may impact neighboring triad(s), which can then affect the meso- and macro-level structure of a network. 
	Looking at the prior body of research on balance collectively, we conclude that for a more comprehensive analysis, balance must be assessed at multiple levels of the network, namely the micro-, meso-, and macro-level. We propose a new methodological framework to ``link micro and macro levels'' \cite{GRANOVETTER1977} in the analysis of social signed networks.
	
	Structural balance has primarily been studied for undirected signed networks \cite{wasserman1994social,leskovec2010signed,chiang2014prediction,aref2015measuring} as opposed to directed and/or signed networks \cite{rapoport1983mathematical,levi1969elementary,sherwin1971introduction}, although directed signed network data have been available since the early days of balance theory \cite{lemann1952group}. Using undirected network models for balance assessment could be a justified choice when the modeled relationships are truly undirected, such as collaboration \cite{aref2020detecting}, or inherently reciprocal, such as bi-lateral alliances \cite{aref2017balance}. However, many real-world relationships are intrinsically directed, such as social preferences \cite{lemann1952group,newcomb_acquaintance_1961}, and not necessarily reciprocated, such as friendship \cite{almaatouq2016you}. Therefore, disregarding directionality  \cite{facchetti_computing_2011}, when it does apply, can jeopardize the validity of calculating network measures \cite{smith2013structural}, including balance assessment \cite{leskovec2010signed,facchetti_computing_2011}.
	
	Several recent studies have implemented distinct methods and measurements to bridge the gap between balance theory and its evaluation on empirical data \cite{maoz2007,estrada_walk-based_2014,Lerner2015,aref2015measuring}. The empirical findings to support the theory seem to be mixed at the first glance. In the context of international relations, Maoz et al.\ \cite{maoz2007} compared the probability of a negative edge between nodes sharing a common enemy and that of a random dyad. They observed the probability of a negative edge to be higher than they had expected and considered this unexpected finding as empirical evidence that contradicts balance theory. Later, Lerner \cite{Lerner2015} explained the same contradictory observation to be an artifact of the method used by Maoz et al.\ \cite{maoz2007}; arguing that by conditioning the probability on the presence of an edge, the empirical observations would become indicative of strong support for balance theory. Similarly, Estrada and Benzi proposed a measure of balance based on closed-walks, and observed that balance in real signed networks is particularly low \cite{estrada_walk-based_2014}; suggesting their observation to be evidence against balance theory. Later, Aref and Wilson \cite{aref2015measuring} showed that such observed low balance is an artifact of the walk-based method used by Estrada and Benzi \cite{estrada_walk-based_2014}, which consistently shows low balance for real networks even in cases where all alternative measures show high balance and provide strong evidence in support of balance theory. These are only two example for the mixed observations on the applicability of balance in real networks which could be traced back to the methods used for evaluating balance on empirical data. This unclear status quo shows that further methodological work is needed to bridge the gap between balance theory and empirical evaluations of balance.
	
	Existing methods in network analysis, such as multilevel modeling \cite{kleinnijenhuis2013adjustment}, relational event models (REMs) \cite{lerner2020free}, and exponential random graph models (ERGMs) \cite{wang2013exponential} have been used to examine balance dynamics at the micro- and macro-level. Kleinnijenhuis and de Nooy \cite{kleinnijenhuis2013adjustment} observed a tendency in political networks towards balance and transitivity, according to which a party's position on an issue is influenced by the positions of their allies and opponents. The authors constructed a cross-nested, multilevel model, which allows for multiple political parties being involved in multiple issues, and found evidence for the micro-level balance effect that impacts a party's position towards an issue. Lerner and Lomi \cite{lerner2020free} constructed an REM to examine micro- and macro-level patterns of positive and negative ties between Wikipedia editors, based on \textit{redoing} and \textit{undoing} events, respectively. The authors found support for all but one pattern of micro-level balance (friend of enemy was not found to be an enemy), and that prominence of balance at the micro-level gave rise to a polarized macro-level structure, where events of the same sign between two editors are likely to occur again. Wang et al.\ \cite{wang2013exponential} conducted a series of experiments to determine the extent to which triadic closure effects, along with other parameters, explain the structural characteristics of a network at the micro-, meso-, and macro-level. Their findings demonstrate structural dependencies between the three levels of analysis, where a change in one level may impact the configuration at another level. We build upon and extend prior findings by examining multiple levels of analysis with respect to structural balance. 
	
	This study makes two primary contributions to address the issues outlined above. First, we propose a multilevel evaluation framework that consists of balance measurement at three fundamental levels of analysis, namely at the (1) triadic (micro), (2) subgroup (meso), and (3) network (macro) levels. Our definition of multilevel evaluation differs from prior literature on multilevel analysis or multilevel models \cite{kleinnijenhuis2013adjustment} in that our proposed framework does not deal with a statistical model with parameters varying in more than one level, neither do we consider the nested nature of the three levels of analysis \cite{lazega2015multilevel,wang2013exponential}. Instead, we evaluate three levels of the network structure separately, and draw inferences from the results about the extent to which balance applies at each of the considered network levels. We then determine if measurements at different levels lead to different conclusions about balance in one and the same the same network. Secondly, to analyze balance, we consider both tie direction and sign. At the micro-level, we examine the balance of semicycles embedded within each triad. At the meso-level, we derive measures of cohesiveness and divisiveness to capture balance within and between subgroups. At the macro-level, we leverage a normalized version of the line index of balance \cite{harary1959measurement} (also known as the normalized frustration index \cite{aref2015measuring,aref2017balance}) to measure partial balance of a whole network. 
	
	
	We analyze 11 real-world networks to determine how balance manifests itself in different social settings (e.g., friendship among students, relationships among philosophers); in some cases with additional network properties (temporality, layering). Temporality is the multiplicity of snapshots of the same network over a period of time. The number of snapshots can be either one (static network), or more than one (temporal network). Layering is the number of different signed relationship types within the same network, which can either be one (single-layer network), or multiple (multilayer network). We divide our datasets into three categories with respect to the additional dimensions: (a) static single-layer (static for brevity) , (b) temporal (multiple snapshots), and (c) multilayer (multiple layers). 
	We aim to answer the following research questions, which are methodological in nature:
	\begin{itemize}
		\item[-] RQ1: How can we measure the balance of a static signed directed network at the micro-, meso-, and macro-level?
		\item[-] RQ2: What insights do we gain from a multilevel evaluation of balance in a static signed directed network?  
		\item[-] RQ3: How is a multilevel evaluation of balance applied to a temporal or multilayer signed directed network? 
	\end{itemize}
	
	Research questions 1 and 2 aim to identify a methodology for understanding balance as a potentially multi-faceted concept. Such a general method has to be applicable to a wide range of networks that differ in social settings and structural properties. The third research question explores the applicability of multilevel balance analysis to temporal and/or multilayer networks. In essence, our work identifies the extent to which networks are balanced depending on level of analysis; thereby bringing together and building on top of insights from previous studies that focus on a single level of analysis for one or more networks.
	Our methodological contribution allows us to reflect on the substantive question about the extent to which social relations show consistency with balance theory \cite{heider1946attitudes, cartwright_structural_1956, harary_measurement_1959, aref2015measuring}.
	
	\subsection*{Our contributions}
	\begin{itemize}
		\item We develop, operationalize and test a novel measure for micro-level balance in signed and directed networks. This measures accounts for edge directionality and is  grounded in theory by bringing together theoretical work on transitivity and balance theory.
		\item We apply balance theory to the meso-level of networks, and provide two new measures for the evaluation of balance that quantify the concepts of cohesiveness and divisiveness of subgroups.    
		\item We re-purpose the frustration index, expand it to incorporate directionality, and develop an optimization model for computing the frustration index of directed and signed networks.
		\item We propose a methodological framework with wide applicability to network structural analysis that allows for considering three levels of analysis to improve the comprehensiveness of evaluating signed and directed networks.
		\item Our analysis of empirical network data demonstrates that the relevance and suitability of our proposed framework not only for static networks, but also for directed and signed network data with additional features (temporal and multilayer). 
	\end{itemize}
	
	\section*{Data Description} \label{sec:data}
	One key purposes of this study is to present a single, general methodology for analyzing signed directed networks. Therefore, we demonstrate the application of our proposed method by analyzing 11 networks that contain signed and directed ties, and represent distinct social contexts. Table \ref{tab:Tab1} provides details and descriptive information for each dataset considered herein. Further details are provided in the Supplementary Information. We analyze eight static networks of different size, ranging from dozens to tens of thousands of edges. An example of a static network is \textit{Reddit}, which denotes positive or negative sentiment (edge type) of content shared between online accounts and captured at one point in time. Furthermore, we study two temporal networks, namely \textit{Sampson's monastery affect} data \cite{sampson1969novitiate} collected over three time periods (T2, T3, T4), and \textit{Newcomb's fraternity network}\cite{newcomb_acquaintance_1961}, which entails 15 snapshots. Our version of the Sampson's network data contain one type or layer of relationship and two possible values (positive or negative) of edges among 18 monks. Newcomb's fraternity network contains one layer of relationship with two possible values (positive or negative) of edges among 17 fraternity brothers living in a shared residence \cite{newcomb_acquaintance_1961}. 
	Finally, we examine one multilayer social network, \textit{Collins' philosophers network} \cite{collins2009sociology}, which consists of one snapshot of two types of relationships (master-pupil and acquaintanceship) between philosophers from different schools of thoughts. 
	To the extent of our knowledge, network data and studies on real-world directed signed networks that are temporal and multilayer are limited. All data used in this study are publicly available under a CC BY 4.0 license in a FigShare data repository \cite{figshare2020directed-signed-graphs}.
	
	\begin{table*}[htb]
		\centering
		\caption{List of directed signed networks datasets used in our study}
		\label{tab:Tab1}
		\resizebox{\textwidth}{!}{%
			\begin{tabular}{|l|c|c|c|c|l|}
				\hline
				\multicolumn{1}{|c|}{\multirow{2}{*}{\textbf{Dataset}}} &  \multirow{2}{*}{\textbf{\begin{tabular}[c]{@{}l@{}}$n$ (\# of nodes)\end{tabular}}} & \multirow{2}{*}{\textbf{\begin{tabular}[c]{@{}c@{}}$m$ (\# of edges)\\ $(m^+,m^-)$\end{tabular}}} & \multicolumn{2}{c|}{\textbf{\begin{tabular}[c]{@{}c@{}}Type of Network\end{tabular}}} & \multirow{2}{*}{\textbf{\begin{tabular}[c]{@{}c@{}}Description \end{tabular}}} \\ \cline{4-5} 
				
				\multicolumn{1}{|c|}{} &  &  &
				\multicolumn{1}{l|}{\textbf{\# of Snapshots}} & \multicolumn{1}{l|}{\textbf{\# of Layers}} & \\ \cline{1-6} \bottomrule
				
				Reddit \cite{kumar2018community} & 18,313 & \begin{tabular}[c]{@{}c@{}}120,792 \\(111,891, 8,901)\end{tabular}& 1 & 1 &\begin{tabular}[c]{@{}l@{}} Represents connections between users of two subreddits from \\ Jan 2014 to April 2017. Collected online. \end{tabular} \\\hline
				
				\begin{tabular}[c]{@{}l@{}}Wikipedia election \cite{west2014exploiting} \end{tabular} & 7,118 & \begin{tabular}[c]{@{}c@{}}103,675 \\(81,318, 22,357)\end{tabular}&1 & 1&  \begin{tabular}[c]{@{}l@{}} Contains approval and disapproval votes for electing \\ admins in Wikipedia from 2003 to 2013. Collected online. \end{tabular} \\\hline
				
				\begin{tabular}[c]{@{}l@{}}Bitcoin OTC \cite{kumar2016edge}\end{tabular} & 5,881 & \begin{tabular}[c]{@{}c@{}}35,592 \\(32,029, 3,563)\end{tabular}&1 & 1 &  \begin{tabular}[c]{@{}l@{}}Represents the record of reputation/trust of users on a \\ Bitcoin trading platform. Collected online. \end{tabular} \\ \hline
				
				\begin{tabular}[c]{@{}l@{}}Bitcoin Alpha \cite{kumar2016edge} \end{tabular} & 3,783 & \begin{tabular}[c]{@{}c@{}}24,186 \\(22,650, 1,536)\end{tabular}&1 & 1 &  \begin{tabular}[c]{@{}l@{}}Represents the record of reputation/trust of users on a \\ Bitcoin trading platform. Collected online. \end{tabular} \\ \hline
				
				\begin{tabular}[c]{@{}l@{}}Highland tribes \cite{read_cultures_1954} \end{tabular} & 16 & \begin{tabular}[c]{@{}c@{}}116 \\(58, 58) \end{tabular}& 1 & 1 & \begin{tabular}[c]{@{}l@{}} Represents alliance structure among three tribal groups. \\ Collected offline. \end{tabular} \\ \hline
				
				\begin{tabular}[c]{@{}l@{}}College preferences\\House A \cite{lemann1952group} \end{tabular} & 21 & \begin{tabular}[c]{@{}c@{}}94 \\(51, 43)\end{tabular}& 1 & 1 & \begin{tabular}[c]{@{}l@{}}Preference rankings of a group of girls in an Eastern \\ college. Collected offline. \end{tabular} \\ \hline
				
				\begin{tabular}[c]{@{}l@{}}College preferences\\House B \cite{lemann1952group}\end{tabular} & 17 & \begin{tabular}[c]{@{}c@{}}83 \\(41, 42)\end{tabular}& 1 & 1 & \begin{tabular}[c]{@{}l@{}}Preference rankings of a group of girls in an Eastern \\ college. Collected offline.\end{tabular} \\\hline
				
				\begin{tabular}[c]{@{}l@{}}College preferences\\House C \cite{lemann1952group}\end{tabular} & 20 & \begin{tabular}[c]{@{}c@{}}81 \\(41, 40)\end{tabular}& 1 & 1 & \begin{tabular}[c]{@{}l@{}}Preference rankings of a group of girls in an Eastern \\ college. Collected offline.\end{tabular} \\ \hline
				
				\begin{tabular}[c]{@{}l@{}}Sampson's affect\cite{sampson1969novitiate}\end{tabular} & 18 & \begin{tabular}[c]{@{}c@{}} T2: 104 (55, 49)\\T3: 105 (57, 48)\\T4: 103 (56, 47)\end{tabular} &3 & 1 &  \begin{tabular}[c]{@{}l@{}} Represents social relationships among 18 monk-novitiates \\ over 3 time periods. Collected offline.\end{tabular} \\ \hline
				
				\begin{tabular}[c]{@{}l@{}}Newcomb's fraternity\cite{newcomb_acquaintance_1961} \end{tabular}& 17 & \begin{tabular}[c]{@{}c@{}} All snapshots:\\ 102 (51, 51)\end{tabular}& 15 & 1 &  \begin{tabular}[c]{@{}l@{}} Preference rankings of 17 boys in a pseudo-dormitory \\ over 13 weeks. Collected offline.\end{tabular} \\ \hline
				
				\begin{tabular}[c]{@{}l@{}}Philosophers network\cite{collins2009sociology} \end{tabular} & 855 & \begin{tabular}[c]{@{}c@{}}2,010 \\(1,736, 274)\end{tabular}& 1 & 2 &  \begin{tabular}[c]{@{}l@{}}Acquaintanceship and master-pupil relationships among \\ philosophers recorded by Collins in \cite{collins2009sociology}  Collected offline.\end{tabular} \\ \hline
				
			\end{tabular}
		}
	\end{table*}
	
	\section*{Notations and Basic Definitions}
	\label{s:preliminaries}
	\label{ss:signedgraph}
	We denote a directed signed graph as $G = (V,E,\sigma)$, where $V$ and $E$ are sets of vertices and directed edges respectively, and $\sigma$ is the sign function that maps edges to $\{-1,+1\}$. A signed digraph $G$ contains $|V|=n$ nodes and $|E|=m$ directed edges. The set $E$ of directed edges contains $m^-$ negative edges and $m^+$ positive edges. 
	
	A \emph{triad} in $G$ is a set of three nodes with at least one directed edge between each two of them (could be in either direction). Figure \ref{fig:Fig1} shows 4 triads. Given a triad, if there are 3 edges incident on its nodes such that for every pair of nodes, there is one edge, then those three edges form a \textit{semicycle}. A triad has at least one semicycle but it could also have multiple semicycles. The leftmost triad in Figure \ref{fig:Fig1} has one semicycle while the rightmost triad has eight semicycles.
	If the binary relation $\mathcal{R}$ that defines edges $A\mathcal{R}B \leftrightarrow (A,B)\in E$ is transitive over the set of a semicycle's edges (i.e.\ $A\mathcal{R}B \ \& \ B\mathcal{R}C \rightarrow A\mathcal{R}C$), the semicycle is called a \textit{transitive semicycle}. A transitive semicycle is balanced (unbalanced) if and only if the product of the signs on its edges is positive (negative). 
	A signed digraph is balanced if and only if its set of nodes can be partitioned into two groups such that all positive edges are within each group and all negative edges are between the groups.
	
	\section*{Multilevel Evaluation of Balance}
	In this section, we discuss our proposed multilevel evaluation framework which involves measuring balance at the micro-, meso-, and macro-level.  
	
	\subsection*{Measuring balance at the micro-level}
\begin{figure}[t]
		\centering
		\centerline{\includegraphics[width=0.8\linewidth]{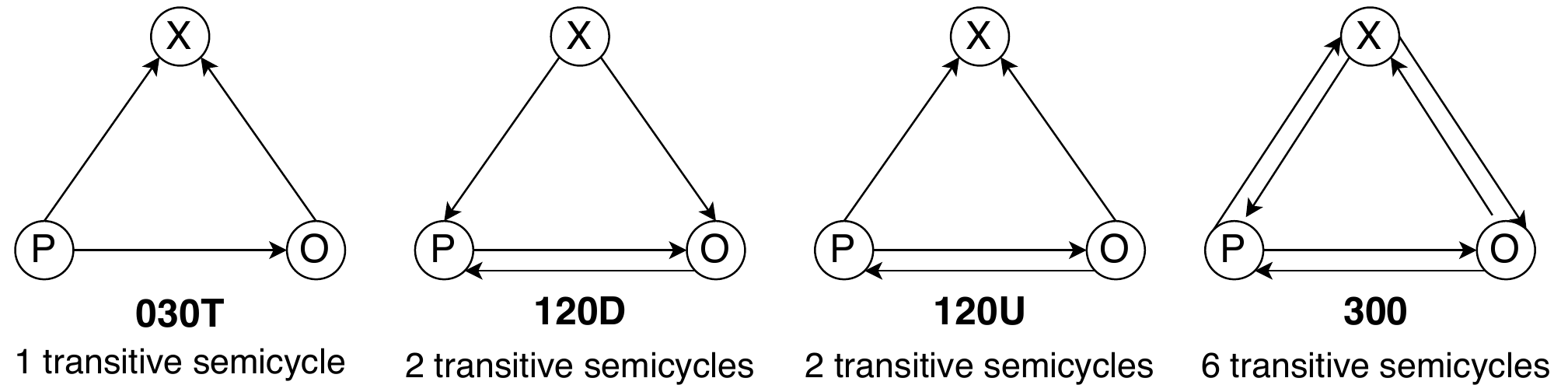}}
		\vspace{2mm}
		\caption{Triads in the triad census \cite{holland1970method} with only transitive semicycles. Signs of edges (not shown in the figure) can either be positive or negative. Triad types are labeled based on the number of mutual (first digit), asymmetric (second digit), and null (third digit) dyads, and an additional letter for direction (T:transitive, D:down, U:up). See \cite{holland1970method} for more details about nomenclature for the triad census.}
		\label{fig:Fig1}
	\end{figure}

	To evaluate balance in a signed network, the most common method is to quantify balance per triads \cite{cartwright_structural_1956,rapoport1983mathematical,johnsen1986structure,terzi_spectral_2011}. This step is usually followed by adding up and comparing frequencies or ratios of balanced versus unbalanced triads, with the implicit assumption being that this aggregation represents a network's overall balance. The majority of studies does not consider edge directionality when calculating triadic balance. 
	In real-world social networks with positive and negative relationships, ties are not necessarily reciprocated. For instance, A might perceive B as a friend, but B is neutral towards A, which can be formulated as $(A,B) \in E^+, (B,A) \notin E$ using a signed digraph notation. Another example would be A trusting B, but B distrusting A, which can be formulated as $(A,B) \in E^+, (B,A) \in E^-$. Undirected signed networks are incapable of modeling such basic cases, leading to not incorporating these situations in network models \cite{facchetti_computing_2011} or disregarding all unreciprocated edges for analysis \cite{diesner2015little,leskovec2010signed}. This fundamental flaw is resolved by using signed digraphs, which allows for creating a more flexible and comprehensive network model. Addressing this problem requires considering edge directionality in the measurement of balance. Our unit of analysis for the micro-level evaluation of balance is a transitive semicycle.
	We only evaluate triads in which all semicycles are transitive (to which we refer to as transitive triads). Four types of triads (as in the triad census \cite{holland1970method}: `030T', `120D', `120U', and `300') are transitive (illustrated in Fig.\ \ref{fig:Fig1}). A transitive triad is balanced if all of its semicycles are balanced. 
	For our analysis, we use the \textit{NetworkX} library in \textit{Python} to assess the balance of triads and obtain $T(G)$ as the fraction of balanced transitive triads over all transitive triads.

	Evaluating balance solely at the micro-level is common practice, which rests on the assumption that aggregating triad-level balance is sufficient to determine network-level balance. 
	Also, measuring balance at the triad level does not consider how configurations within triads influence neighboring nodes and edges as well as broader areas of the network. Based on prior literature, there are structural configurations beyond the triad, such as longer cycles, that contribute to balance or lack thereof \cite{harary1959measurement,doreian_partitioning_1996,bonacich_introduction_2012,estrada_walk-based_2014,aref2015measuring,aref2017balance}, which shows that simply aggregating balance scores from the micro-level might not capture other structural features.
	To mitigate the limitations resulting from single-level evaluation, we propose and apply complementary methods to evaluate meso- and macro-level balance as parts of a comprehensive multilevel evaluation framework.
	
	\subsection*{Measuring balance at the meso-level} 
	To evaluate meso-level balance, vertices of a network can be partitioned into two mutually antagonist but internally solidary subgroups \cite{zaslavsky_balanced_1987, zaslavsky_mathematical_2012,aref2017computing,aref2016exact}. An internally solidary subgroup means that there are only positive edges within a subgroup. Two internally solidary subgroups are mutually antagonistic when they are connected by only negative edges. This approach returns the minimum number of negative edges within subgroups and positive edges across subgroups across all possible ways of partitioning nodes into two subsets.
	We refer to solidarity within subgroups as cohesiveness, and to antagonism between subgroups as divisiveness within a network. To illustrate our approach for quantifying these two new measures, we start with an extreme yet illustrative example: A balanced network that contains only negative edges has an extreme amount of cohesiveness (because all edges within its two subgroups are positive) and an extreme amount of divisiveness (because all edges between its two subgroups are negative). We quantify cohesiveness and divisiveness through the deviation from this extreme case.
	
	Using a signed digraph $G=(V,E,\sigma)$ as input, the set of vertices, $V$, can be partitioned based on $P=\{X,V\setminus X\}$ into the two subgroups $X$ and $V\setminus X$. Given partition $P=\{X,V\setminus X\}$, edges that cross the subgroups are \textit{external} edges that belong to $E^e_P=\{(i,j)\in E |i\in X, j\notin X \ \text{or} \ i\notin X, j\in X\}$. Edges that do not cross the subgroups are \textit{internal} edges that belong to $E^i_P=\{(i,j)\in E |i,j\in X \ \text{or} \ i,j\notin X\}$. 
	We measure cohesiveness (divisiveness) of a partition $P$ by only looking at the signs of its internal (external) edges. We quantify the cohesiveness of a given partition $P$ using the fraction of its positive internal edges to all internal edges $C(P)={| E^i_P \cap E^+|}/{|E^i_P|}$. Similarly, we quantify the divisiveness of partition $P$ as the fraction of its negative external edges to all external edges $D(P)={| E^e_P \cap E^-|}/{|E^e_P|}$. 
	We compute cohesiveness and divisiveness using $P^*$, which is the best fitting bi-partition of nodes as explained further in the next subsection, and the bi-partition is also connected to our proposed macro-level analysis.
	Our proposed measures of cohesiveness and divisiveness are consistent with prior social networks literature, especially with the concepts of ranked clusterability \cite{wasserman1994social}, partitioning nodes via blockmodeling \cite{doreian_partitioning_1996}, in-group attraction and out-group repulsion mechanisms \cite{stadtfeld2020emergence}, and intra- and inter-group conflicts in small groups \cite{crawford2013intra-inter,cohen2008war}, as well as sociological literature on faultline theory \cite{lau1998faultline-theory}. The theoretical underpinnings of our proposed methodology are discussed further in the Supplementary Information.
	
	\subsection*{Measuring balance at the macro-level}  
	The line index of balance, denoted as $L(G)$ and also referred to as the \textit{frustration index} \cite{zaslavsky_balanced_1987,aref2017balance} and \textit{global balance} \cite{facchetti_computing_2011}, is defined as the minimum number of edges whose removal leads to balance. These edges can be thought of as sources of tension in this approach. While the historical roots of the frustration index go back to the 1950's \cite{abelson_symbolic_1958,harary1959measurement}, this approach only started to receive major attention in recent years \cite{iacono_determining_2010,facchetti_computing_2011,aref2015measuring,aref2017balance,aref2020detecting}. This might be due to the computational complexity of obtaining this index exactly, which is a NP-hard problem \cite{huffner_separator-based_2010}. While even approximating this measure has been difficult \cite{iacono_determining_2010,facchetti_computing_2011}, recent developments have enabled the exact and efficient computation of $L(G)$ for graphs with up to $10^5$ edges \cite{aref2017computing,aref2016exact}. To the best of our knowledge, the frustration index has not been previously applied to or computed for directed graphs. As a technical contribution, we have developed the first exact method for computing this index for directed signed graphs by building on top of recently proposed optimization models \cite{aref2016exact,aref2020detecting}.
	
	Frustration of an edge depends on how the edge resides with respect to the partition $P=\{X,V\setminus X\}$ that is applied to $V$. Positive edges with endpoints in different subsets and negative edges with endpoints in same subset are frustrated edges under $P$. The frustration index offers a top-down evaluation mechanism for assessing partial balance by providing an optimal partition $P^*$. The optimal partition $P^*$ minimizes the number of frustrated edges and is therefore the best fitting partition of nodes into two mutually antagonistic and internally solidary groups. A simple normalization of $L(G)$ using a line index upper bound (which equals a half of the edge count, $m/2$ \cite{aref2017computing}) leads to the normalized line index $F(G)=1-2L(G)/m$ \cite{aref2015measuring}. The normalized line index provides values within the unit interval such that large values represent higher partial balance and therefore higher consistency of a network with balance theory at the macro-level. More details on this measure and the optimization model we use for this computation are provided in the Supplementary Information. We solve the optimization model using \textit{Gurobi} solver \cite{gurobi} (version 9.0) in \textit{Python}. For large networks, we follow the two-step method presented in \cite{aref2020detecting}, which involves computing a lower bound for the frustration index before solving the optimization model.
	
	\section*{Results and Discussions}
	\subsection*{Measuring balance of static networks at three levels}
	
	Our first two research questions ask to what degree our impression of the balance of a network depends on the chosen level of analysis (micro, meso, or macro). To answer these questions, we quantify balance for each level separately, and then base our interpretation of the balance of the network on the results across levels. 
	
	To measure balance of static signed directed networks at the micro-level, we use $T(G)$ the proportion of balanced triad in a network among all transitive triads. 
	Table \ref{tab:Tab2} shows that triad-level balance values are high across all static networks with an average of 0.78 (Min = 0.52, Max = 0.90, SD = 0.12), except for the \textit{College House B}, where only 52\% of triads satisfy the semicycle balance property. 
	Our results are consistent with the central tenet of balance theory, which states that networks strive towards stability in their triadic configurations, which then leads to high proportions of balanced triads and reduced tension \cite{hummon_dynamics_2003}. Note that a micro-level measurement of balance based on triads 
	may fall short in sparse networks where triads are infrequent and the clustering coefficient (the fraction of closed triples to all triples) is low \cite{aref2015measuring}. Therefore, values of $T(G)$ for \textit{Reddit}, \textit{Wikipedia}, \textit{Bitcoin Alpha}, and \textit{Bitcoin OTC} are more suitable to be interpreted as measurements of triads per se as opposed to that of the overall structure sampled through the triads.
	
	\begin{table}[!b] 
		\caption{Balance results for static networks}
		\label{tab:Tab2}
		\centering
		\resizebox{\textwidth}{!}{%
			\begin{tabular}{|l|c|c|c|c|c|c|c|}
				\hline
				\multicolumn{1}{|c|}{\multirow{2}{*}{\textbf{Network}}} & {\multirow{2}{*}{\textbf{Density}}} & \multirow{2}{*}{\textbf{\begin{tabular}[c]{@{}c@{}} Clustering \\ Coefficient \end{tabular}}} & \multirow{2}{*}{\textbf{\begin{tabular}[c]{@{}l@{}}Triad Level \\ Balance $T(G)$\end{tabular}}} & \multicolumn{2}{c|}{\textbf{\begin{tabular}[c]{@{}c@{}}Subgroup Level Balance\\\end{tabular}}} & \multirow{2}{*}{\textbf{\begin{tabular}[c]{@{}c@{}}Network Level \\Balance $F(G)$\end{tabular}}} & \multirow{2}{*}{\textbf{\begin{tabular}[c]{@{}c@{}}Triad Census \\ (Transitive and Balanced) \end{tabular}}} \\ \cline{5-6} 
				
				\multicolumn{1}{|c|}{} & & &  & 
				\multicolumn{1}{l|}{\textbf{Cohesiveness $C(P^*)$}} & \multicolumn{1}{l|}{\textbf{Divisiveness $D(P^*)$}} &  &\\ \cline{1-8} \bottomrule
				Reddit & 3.60E-4 & 6.30E-02 & 0.704 & 0.936 & 0.096 & 0.859 & `300': 3.3\%; `120D': 10.3\%; \\ & & & & & & & `120U': 9.7\%; `030T': 47.1\%\\ \hline
				Wikipedia election & 2.00E-03 & 5.30E-02 & 0.751 & 0.869 & 0.765 & 0.710 & `300': 0.27\%; `120D': 6.1\%; \\ & & & & & & & `120U': 7.2\%; `030T': 61.5\%\\ \hline
				Bitcoin OTC & 1.00E-03 & 4.50E-02 & 0.866 & 0.960 & 0.871 & 0.908 & `300': 53.4\%; `120D': 7.7\%; \\ & & & & & & &`120U': 10.5\%; `030T': 15.0\%\\ \hline
				Bitcoin Alpha & 1.70E-03 & 6.40E-02 & 0.845 & 0.960 & 0.781 & 0.909 & `300': 61.4\%; `120D': 6.5\%; \\ & & & & & & & `120U': 10.8\%; `030T': 5.8\%\\ \hline
				Highland Tribes & 0.483 & 0.527 & 0.870 & 0.806 & 1.000 & 0.759 & `300': 87\%\\ \hline
				\begin{tabular}[c]{@{}l@{}}College - House A\end{tabular} & 0.224 & 0.392 & 0.807 & 0.793 & 0.861 & 0.638 & `300': 7.0\%; `120D': 19.3\%; \\ & & & & & & &`120U': 22.8\%; `030T': 31.6\\ \hline
				\begin{tabular}[c]{@{}l@{}}College - House B\end{tabular} & 0.305 & 0.398 & 0.522 & 0.739 & 0.811 & 0.542 & `300': 6.5\%; `120D': 2.2\%; \\ & & & & & & & `120U': 23.9\%; `030T': 19.6\%\\ \hline
				\begin{tabular}[c]{@{}l@{}}College - House C\end{tabular} & 0.213 & 0.271 & 0.896 & 0.909 & 0.973 & 0.877 & `300': 3.4\%; `120D': 27.6\%; \\ & & & & & & & `120U': 10.3\%; `030T': 48.3\%\\ \hline
			\end{tabular}%
		}
	\end{table}
	
\begin{figure}[t]
		\vspace{-3mm}
		\centering
		\centerline{\includegraphics[width=0.35\textwidth]{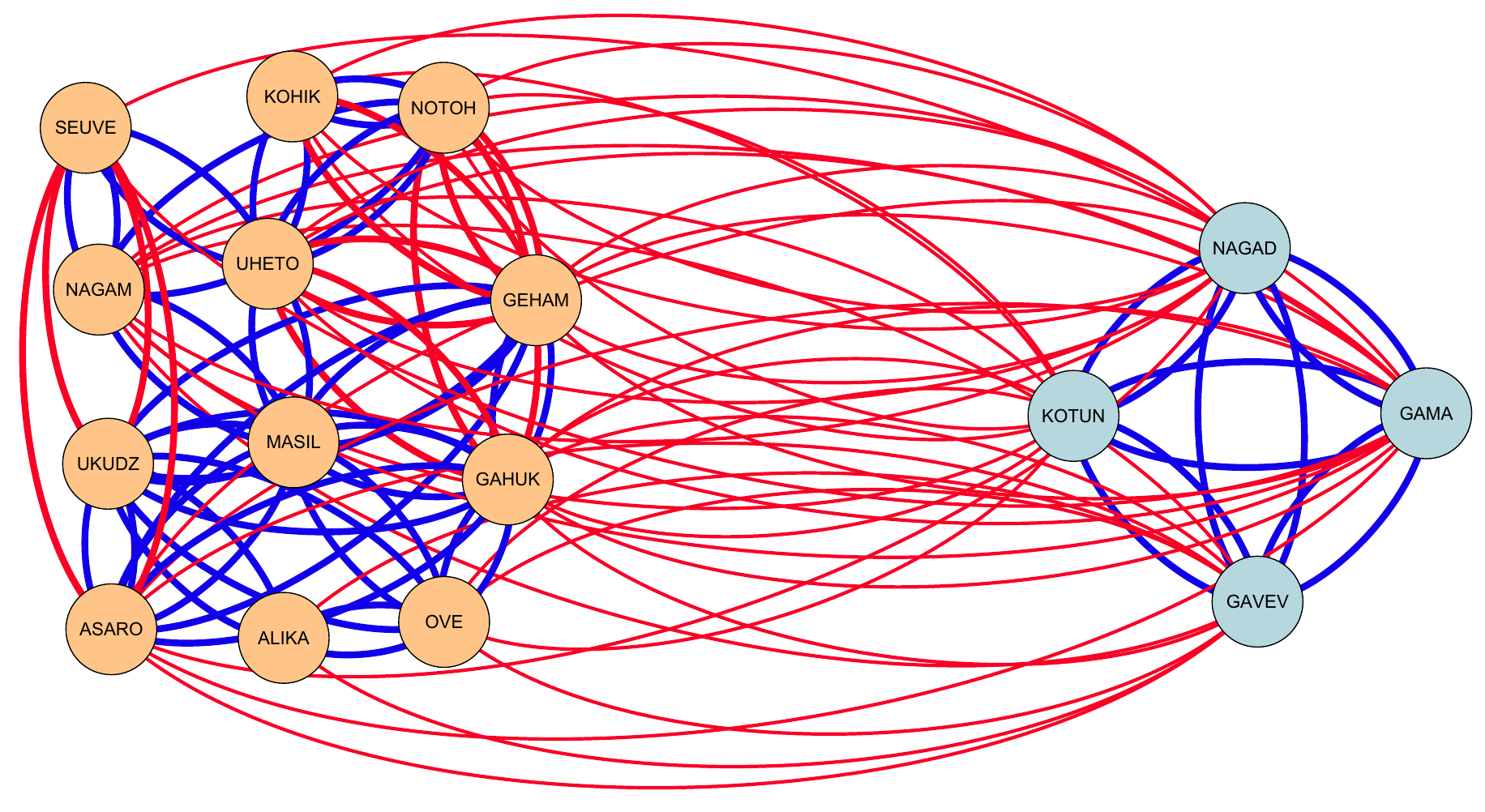}}
		\caption{Visualization of optimal partition for Highland tribes network (low cohesiveness, high divisiveness). Direction of arcs is clockwise. Blue arcs are positive and red arcs are negative.}
		\label{fig:Fig2}
	\end{figure}

	To evaluate balance of static, signed and directed networks at the meso-level, we compute our proposed measures of cohesiveness (intra-group solidarity) and divisiveness (inter-group antagonism). The numerical results for both metrics are provided in Table \ref{tab:Tab2}. We observe cohesiveness to be high with an average of 0.87 (Min = 0.74, Max = 0.96, SD = 0.08). Divisiveness is also high (except for the \textit{Reddit} network), with an average of 0.77 (Min = 0.10, Max = 1.00, SD = 0.29). The meso-level values for most networks seem to indicate a positive association between the two measures. 
	In other words, we observe high meso-level balance in networks where nodes within the same subgroup are positively tied to their subgroup members, while at the same time, they are negatively tied to members of the other subgroup. This observation is consistent with literature on small-group conflicts, where strong in-group identity \cite{ashleigh2001trust} and weak out-group identity \cite{cohen2008war} signify subgroup boundaries \cite{crawford2013intra-inter}. 
	The \textit{Reddit} network deviates from this general pattern as it shows high cohesiveness ($0.936$) and low divisiveness ($0.096$). While one would expect negative edges to dominate the ties between groups, the visualization (in the Supplementary Information) shows a prominence of positive edges (in blue) in general and between groups, which suggests the existence of one cohesive community rather than two divided subgroups for this particular network.
	Another deviation from the generally observed patterns is seen in the \textit{Highland tribes} network, which has a lower value for cohesiveness compared to its high divisiveness. The visualization (Fig.\ \ref{fig:Fig2}) of this particular network shows the complete dominance of negative edges between subgroups, which explains the extreme divisiveness value ($D(P^*)$=1.00). However, negative edges are also present in one of the subgroup (13.5\% of all edges in the left subgroup). In other words, while two subgroups are clearly divided, there is also some division within one of the subgroups, which influences the overall cohesiveness of subgroups. A closer look at Fig.\ \ref{fig:Fig2} shows that there are only two tribal alliances, `Masil'-`Uheto' and `Masil'-`Nagam', that keep the left subgroup together, while seven pair of tribes in the left subgroup are mutually antagonistic. This lack of cohesion is accounted for by our measurement and consequently impacts the cohesiveness value.

	To measure macro-level balance of static signed directed networks, we compute the normalized line index of balance. We find this index to be high with an average of 0.78 (Min = 0.54, Max = 0.91, SD = 0.14) for the static networks. The \textit{College House B} network has the lowest normalized line index ($F(G) = 0.54$). Moreover, our results show that the measured values for \textit{College House B} is the lowest at both the micro- and macro-level (but not at the meso-level). In other words, low proportion of balanced triads (52.2\%) and a low normalized line index both suggest that tension is present in the network, and that we capture this effect consistently with different balance assessment methods. 
	The \textit{Bitcoin Alpha} network has the highest macro-level ($F(G) = 0.91$) and meso-level cohesiveness ($C(P^*) = 0.960$), and the fourth highest cohesiveness based on the the micro-level ($T(G) = 0.845$). This network shows a profile of balance across different levels, such that balance is reflected in high proportions of balanced triads, high cohesiveness within subgroups and medium-high divisiveness between subgroups, and relatively few frustrated edges in its optimal partitioning. Similar to the prior example, we conclude that this network is fairly balanced, and our metrics consistently suggest this conclusion. Our method offers a profile of balance for distinguishing input networks with respect to balance instead of a single value.
	
	\subsection*{Insights from multilevel evaluation of balance in static networks}
	Our second research question asks about insight gained from applying our proposed multilevel evaluation framework to static signed directed networks. We observe similarities between micro-level balance and macro-level balance when networks are dense and have high clustering coefficients. Consistent with observations in a previous study (see Fig. 2 in \cite{aref2020detecting}), we find that aggregating micro-level balance results could represent macro-level balance when a network is densely connected and is primarily consisted of closed triads. On the other hand, for sparse networks, aggregating micro-level balance would not necessarily lead to similar results as conducting a macro-level evaluation since such networks cannot be reconstructed through the aggregation of their triads. 
	
	The \textit{Reddit} network exemplifies this situation, with a low density of $3.6e-04$ and a low clustering coefficient of $6.30e-02$. While this network's micro-level balance is $0.704$, its numerous positive edges have led to a high cohesiveness of $0.936$, and this effect then translates into a macro-level balance of $0.859$. With lower intensity. A similar situation is observed for \textit{Bitcoin Alpha} and \textit{Bitcoin OTC} (both visualized in the Supplementary Information), which are also characterized by low density ($1.70e-03$ and $1.00e-03$, respectively) and low clustering coefficients ($6.4e-02$ and $4.5e-02$, respectively). The possibility of large difference between balance in sparse networks when measured at the micro-level versus macro-level has been also observed in prior work (see p.23 in \cite{aref2015measuring}).
	For the \textit{Wikipedia} network (which also has a low density and low clustering coefficient), balance measures at the micro- and macro-level are similar. The consistency in balance values of the results above suggests that the sources of tension are well-represented in the triads of the network as well as in the overall macro-level structure. While there are cases where the two measurements match, balance in micro and macro levels are not generally the same property measured at different levels. 
	
	
	\subsection*{Multilevel balance in temporal and multilayer networks}
	Our third research question is about the generalizability of our balance evaluation methodology to temporal and multilayer networks. To demonstrate the generalizability of our proposed framework, we apply it to two temporal networks (\textit{Sampson's monastery} and \textit{Newcomb's fraternity} and one multilayer network (\textit{Collins philosophers}).
	
\begin{figure}[ht]
		\centering
		{\includegraphics[width=0.8\linewidth]{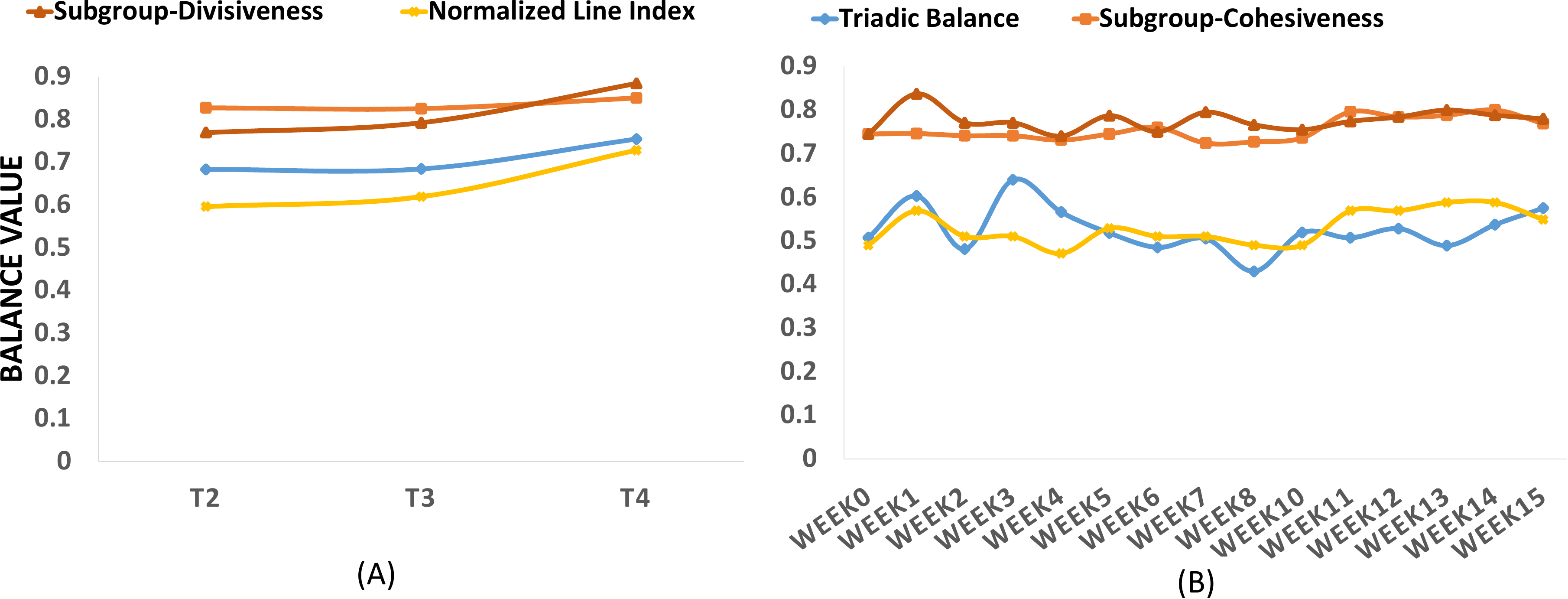}}
		\caption{Multilevel balance values over time for Sampson's affect (A) and Newcomb's fraternity (B) networks}\label{fig:Fig3}
	\end{figure}

	Measurements of balance for the two temporal networks, \textit{Sampson's monastery} and \textit{Newcomb's fraternity}, are shown in Fig.\ \ref{fig:Fig3}. For \textit{Sampson's monastery}, triadic balance (blue line) has an average value of 0.71. The average of cohesiveness (orange line) is 0.83, while divisiveness (red line) has an average of 0.82. The normalized line index (yellow line) has an average value of 0.65. 
	The overall trend is that measurements in all levels show that balance increases over time; supporting the basic premise of balance theory that networks may move towards balance \cite{heider1946attitudes,cartwright_structural_1956,doreian1996brief,aref2017balance}.

\begin{figure}[ht]
		\centering
		{\includegraphics[width=\linewidth]{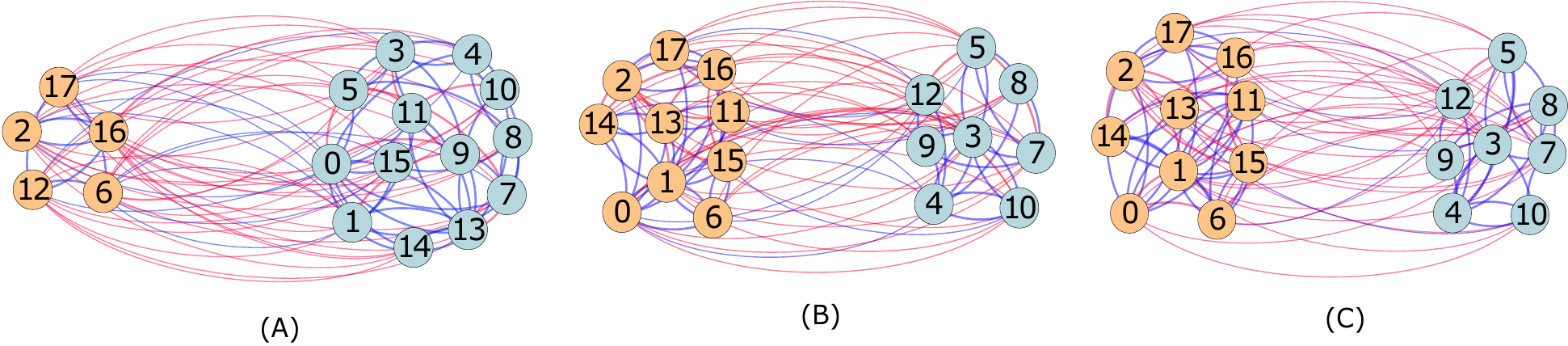}}
		\caption{Visualization of optimal partitions for Sampson's affect network over three snapshots, with (A) for T2, (B) for T3, and (C) for T4}
		\label{fig:Fig4}
	\end{figure}

	Using optimal partitions, we also examine subgroup membership over time for the \textit{Sampson's monastery} network to analyze the changes in how nodes form subgroups. We find that membership in the two subgroups changes from T2 to T3, and remains unchanged from T3 to T4 (shown in Fig.\ \ref{fig:Fig4}). Interpreting these results while considering the groups in Sampson's study \cite{sampson1969novitiate} tells us that at T2, there is a smaller (blue) group of ``outcasts'' \cite{sampson1969novitiate,freeman2017research}, and a larger (yellow) group that consists of the dominant ``young turks'' and the ``loyal oppositions'' led by nodes 1 and 3, respectively. At T3, ``outcasts'' and ``young turks'' form one subgroup (yellow), while the ``loyal oppositions'' and ``waverers'' (those who did not identify with any faction) are in the other subgroup (blue). Interestingly, the optimal partitions remain unchanged in T4. According to Sampson \cite{sampson1969novitiate}, the increase in polarization could be linked to the eventual disintegration of the monastery after T4, when some monks voluntarily left, while others (nodes 1,2,16,17) were asked to leave.
	
	Balance values for \textit{Newcomb's fraternity} network are shown in Fig.\ \ref{fig:Fig3} (right). Micro-level values are, on average, 0.53 (Min = 0.43, Max = 0.64, SD = 0.05). Cohesiveness has an average of 0.75 (Min = 0.72, Max = 0.80, SD = 0.025), and divisiveness has an average of 0.77 (Min = 0.74, Max = 0.83, SD = 0.024). The normalized line index for macro-level balance has an average value of 0.53 (Min = 0.47, Max = 0.59, SD = 0.04). Interestingly, the low values for triadic balance and the normalized line index show that this network has an overall low level of partial balance when assessed at both the micro and macro level. Hence, this network's ties are more in conflict with one another compared to other networks considered in this paper. However, the process of inferring signed ties could impact balance values. 
	
	Similar to the patterns observed for \textit{Sampson's monastery} network, the \textit{Newcomb's fraternity} network has higher balance values at the meso-level and lower values at the micro- and macro-level. In this network, we do not see a clear temporal trend of increasing balance regardless of level of analysis (a time series decomposition did not show a trend). Instead, we see balance measurements oscillate around a narrow range of values over time, which is in contrast to observations of \cite{doreian1996brief} about another model of Newcomb's fraternity data later criticized in \cite{doreian2001pre}. Our observation is consistent with previous studies suggesting that, for some networks, there is no clear tendency to move towards balance over time \cite{belaza2017}. There is more variation over time in triadic balance and divisiveness compared to cohesiveness and normalized line index (as shown in Figure \ref{fig:Fig3}, B). Such differences suggest the presence of social mechanisms that only influence balance at certain levels, as captured by each different measurement of balance. In week 3, for example, there is a notable increase in micro-level balance, while balance at other levels remains the same. A closer look at the types of triads reveals a substantial increase in balanced `030T' (which has 1 transitive semicycle) triads, and a substantial decrease in unbalanced `120D' (which has 2 transitive semicycles) triads. One reason could be that, according to Newcomb \cite{newcomb_acquaintance_1961}, friendships started to appear from week 3 onward. Friendship is a social precondition for reciprocity, transitivity, and balance at the dyadic and triadic levels, which may explain the observed increase in triadic balance in week 3. 
	
	Measurements of balance for the \textit{Philosophers} multilayer network (visualized in Fig.\ S2) show changes in values depending on whether the two layers of the network (acquaintanceship and master-pupil) are analyzed separately or jointly. At the macro-level for instance, the optimal partitions obtained separately for layers of the network are associated with 4 and 6 frustrated edges, while the optimal partition for the whole network (considering both layers) has 60 frustrated edges. This suggests that most sources of tension at the macro-level operate across layers. Specifically for this network, the two layers are interdependent in that three philosophers may be connected through a mix of both types of ties: master-pupil and acquaintanceship. When we analyze the two layers independently, we observe triadic balance values of 0.92 for acquaintance relations, and 0.95 for master-pupil relations, which largely indicate the absence of tension. When the two layers are combined into one multilayer network, the triadic balance value decreases to 0.80. We observe a similar pattern of lower intensity of balance when analyzing the two types of relationships jointly instead of separately at the subgroup and network levels. Meso-level cohesiveness and divisiveness decrease from 0.99 and 0.98 to 0.97 and 0.93, respectively, and the normalized line index decreases from 0.98 to 0.94, when the two types of ties are considered simultaneously. These results indicate that the sources of tension involve groups of philosophers that were connected by a mixture of acquaintance and master-pupil relations and therefore manifest across layers of the network, such that they cannot be detected in either of the layers.  
	Collins discussed that while master-pupil relationships may seem tension-free, in reality, how ``ideas are created has always been a discussion among oppositions'' (p.1 in \cite{collins2009sociology}). 
	
	Methodologically, our observations from analyzing two temporal networks and one multilayer network show that proposed multilevel framework not only allows for analyzing balance as a structural property across three levels, but also for capturing dynamics of balance over time and across different layers of relations. Substantively, the observed temporal dynamics could show a network moving towards higher polarization (\textit{Sampson's network}) or a network without a monotone trend, oscillating over time within a short range of balance values (\textit{Newcomb's fraternity}). Our results for the multilayer network shows that the co-existence of multiple types of signed ties may impact balance of a multilayer network in a more profound manner compared to the individual impact of each type of relation in its respective single-layer network. 
	
	\subsection*{Methodological findings and implications}
	Several studies have compared triadic level balance with estimates for the line index of balance, and found that the line index (as a measure of unbalance) correlates with other measures of unbalance, such as the proportion of unbalanced triads \cite{doreian_reflections_2017, patrick_doreian_structural_2015}. Using normalized versions of triadic balance and the line index, high correlation is observed for networks with high density \cite{aref2015measuring,aref2020detecting}, while measurements for sparse networks usually do not match \cite{aref2015measuring}. However, the literature does not clarify whether such correlations are due to similarities in measurement or structural mechanisms in networks that yield similar balance values at different levels. Here, we briefly examine bivariate associations, if any, between triad-level, subgroup-level, and network-level balance and their associations with other network properties. 
	
	For the static networks, we find a positive and statistically significant correlation between micro-level balance and macro-level balance (Pearson's $r=0.697, p<0.05$) (see the Supplementary Information for theoretically proven connections between the two measures). In addition, there is a positive correlation between meso-level cohesiveness and macro-level balance ($r=0.944, p<0.001$) (these two measures are both related to the optimal partition $P^*$). 
	In the Newcomb's fraternity temporal networks, we find positive and statistically significant correlations between macro-level balance and both meso-level balance measures (macro-level and cohesiveness: $r=0.849, p<0.001$; macro-level and divisiveness: $r=0.717, p<0.01$). These results demonstrate that the correlations among the balance measures could be complex, non-linear, and may depend on the network's structural characteristics. In contrast to prior studies that suggest that micro-level balance induces macro-level balance \cite{chiang2014prediction,srinivasan2011local}, our methodological framework deals with each balance measurement independently, and we do not claim any causality between these concepts. Rather, we assume that there are different theoretical and social mechanisms that regulate balance at each level of analysis \cite{monge2003theories,lerner2020free}. Further studies are needed to substantiate any potential causal links. The implication of our framework at this stage is to facilitate the evaluation of balance of directed signed networks at three levels of analysis with respect to the social processes involved at each level. Given the mixed observations and inconsistencies in empirical evaluations of balance, we consider this implication as a fundamental step towards a multilevel and multitheoretical framework \cite{monge2003theories} for explaining the co-existence of distinct balance processes at the micro-, meso-, and macro-level.
	
	\section*{Conclusion}
	In this study, we developed and applied a multilevel framework for the computational analysis of balance in signed directed networks. Our analysis of a wide variety of networks (including temporal and multilayer networks) representing various social settings (from college students and Wikipedia editors to philosophers and Bitcoin traders) shows that balance presents differently across multiple levels of the networks; leading to different profiles of social structures in triads, subgroups, and the whole network. In most cases, we observe relatively high values of balance across the three levels under consideration despite the differences in social setting and types of signed ties. Our study also serves as another confirmation of balance theory, with which the analyzed networks show a partial but considerable consistency. Our comprehensive numerical results show that values of balance at the micro-, meso-, and macro-level may match up to some extent. In the absence of other network dynamics, for which we have not tested, our findings suggest that the underlying mechanisms of avoiding tension and conflict may be reflected in micro-level patterns of balanced transitive semicycles. At a higher level, such patterns form a meso-level of internally cohesive and externally divisive subgroups. These effects eventually gives rise to a macro-level polarization, where only a relatively few edges are positioned inconsistently with respect to the assertions of balance theory. We also find empirical evidence for the argument that balance at different levels is not just the same social phenomenon measured at different levels, but represents different properties (triadic balance, internal cohesion of subgroups, external division of subgroups, and overall polarization of the network) and should be evaluated independently from one another. We have provided empirical evidence that suggests that these distinct structural properties of signed social networks are inextricably intertwined by a set of fundamental rules of partial balance. These fundamental rules could be expressed at each of the three levels of the networks: (1) In networks partially balance at the micro-level, balanced triads outnumber unbalanced triads. (2) Networks partially balanced at the meso-level have an optimal partition with two internally cohesive and externally divisive subgroups. (3) Networks partially balance at the macro-level are relatively a few number of edges away from balance. 
	
	Our study is an intermediate step that advances our knowledge about the structure of signed social networks. The generalizability of our findings is subject to certain limitations. First, we analyze networks with up to 120,000 edges (upper bound due to the demanding complexity of our exact computational methods). We hope to extend this multilevel framework to an even wider range of social networks before confirming what this study has partially substantiated for the first time. Second, we aim to include more temporal and multilayer networks, and combining the two in future studies in order to generalize our observations with respect to how balance at different levels may change over time and across types of relationships. 
	
	\section*{Materials and Methods}
	We analyze 11 networks that have ties with signs and directionality. These data are publicly available for research purposes. Here, we briefly outline the context and collection of primary data, and how they were processed into network data. Technical details on network data pre-processing are provided in the Supplementary Information.
	
	\subsection*{Static networks} 
	Most of the static networks in our data were previously used for network-related research. Specifically, the \textit{Reddit} dataset \cite{kumar2018community} represents how content with negative sentiment was exchanged between posts from two different communities of users (subreddits). \textit{Wikipedia election} network \cite{west2014exploiting} was collected to study the relationship between sentiment and person-to-person evaluation of leadership and credibility. Hence, the data represent approval or disapproval of voters for deciding on promoting a user to be a Wikipedia administrator. \textit{Bitcoin-OTC} and \textit{Bitcoin-Alpha} were both built from user ratings of trust towards other users on the Bitcoin online trading platforms. The authors \cite{kumar2016edge} used the data to build a prediction model for edge weights using measures of trust between individuals. In contrast to these three datasets, which were collected from online sources, \textit{Highland tribes} and \textit{College Houses A,B,C} data were collected through fieldwork and surveys, respectively. \textit{Highland tribes} \cite{read_cultures_1954} was created through fieldwork that included observations and conversations with political leaders and tribal members of the Central Highlands in New Guinea. The signed network represented political alliances and oppositions among 16 tribes. Network data for College Houses A,B,C \cite{lemann1952group} come from a small group survey on evaluations of other group members (sorority sisters) based on a range of behavioral characteristics. 
	
	\subsection*{Temporal and multilayer networks} 
	Two temporal networks, \textit{Sampson's affect} and \textit{Newcomb's fraternity}, were both collected for qualitative social network analysis. \textit{Sampson's affect} \cite{sampson1969novitiate} was collected via 12 months of fieldwork at an American monastery, where interactions between 18 monks were recorded. The author also created a questionnaire asking each monk to list ``3 brothers who you like the most'' in order to examine the friendship network among different subgroups of monks (young turks, outcasts, loyal oppositions, and waverers). \textit{Newcomb's fraternity} data was created from a 15-week survey of 17 fraternity brothers at the University of Michigan in 1956. The temporal nature of the dataset allowed for in-depth examination of the formation of acquaintanceship and social groups. 
	For the \textit{Philosophers} network data, master-pupil and acquaintanceship ties between philosophers from 800 B.C.E to 1935 C.E were gathered from Randall Collins' seminal book \cite{collins2009sociology} based on close readings of historical texts. Among many findings, through an examination of the ties between different groups of philosophers, the author found that successful philosophers had the most ties to other philosophers, regardless of the signs of ties.
	
	\section*{Availability of Data and Code}
	All network data and code used in this study are made publicly available. Links and descriptions for data and code are provided in the Supplementary Information.


	\section*{Author Contributions}
	S.A., L.D., and R.R. have equally contributed to designing and conducting the research, which primarily involved reviewing the literature, developing and implementing new methods and algorithms, analyzing data and results, and writing the article and the supplementary information; J.D. guided and supervised the research and writing. 
	
	\section*{Acknowledgments}
	The authors are thankful for the assistance with network data pre-processing and results preparation provided by Kehan Li. They wish to acknowledge Wouter de Nooy for providing the data on philosophers networks. They are grateful to Andr\'{e} Grow, Emilio Zagheni, Diego Alburez, Carolina Coimbra, Sofia Gil Clavel, Fanny Kluge, and Ugofilippo Basellini for comments and discussions, which helped to improve this article. There is no funding to be reported for this study.
	
	\section*{Additional Information}
	The authors declare no competing interests.

\clearpage

\noindent
The rest of this document describes all supplementary information for the article ``Multilevel Structural Evaluation of Signed Directed Social Networks based on Balance Theory.''

\section*{Supplementary Information}
Supplementary Text\\
Detailed Materials and Methods\\
Figs. 5 to 10\\
Tables 3 to 4\\
Captions for Movies S1 to S3\\
Captions for Databases S1 to S3\\
References 64-65

\section*{Supplementary Text}

\subsection*{Data availability} 
All network data and numerical results related to this study are publicly available with links provided in this document. The code for the computational analysis including the optimization models used is this study will be made publicly available on the Github repository \url{https://github.com/saref/multilevel-balance} upon publication of the paper.

\subsection*{Additional information about the networks}

In this subsection, we provide detailed descriptions of all the data sources used in the study to facilitate contextualization of the actual real-world system or group that is modeled as a signed directed network. 

The Reddit network represents directed signed ties of communication between two users (belonging to different subreddits) inferred through sentiments of multiple posts and comments they have exchanged.
Each connection (edge) consists of three different attributes: the time, the sentiment of the source post, and the text vector of the source post. The network spans over 2.5 years (Jan 2014 to April 2017) and includes all the public messages shared on Reddit in that time period \cite{kumar2018community}. 

The Wikipedia election data consists of all votes (supporting, neutral, or opposing) in elections for promotion of a user to admin role (when a user requests promotion to adminship. The network includes all votes that were cast from 2003 through May 2013 \cite{west2014exploiting}. 

The Bitcoin OTC and the Bitcoin Alpha datasets both represents trust between Bitcoin traders on two platform called Bitcoin OTC and Bitcoin Alpha. The transactions on these platforms are anonymous. However, to minimize fraudulent behavior and to maintain reputation, members of OTC and Alpha rate others in a scale of -10 (total distrust) to +10 (total trust) in steps of 1 \cite{kumar2016edge}.

The Highland tribes dataset represents alliance structure among tribal groups using Gahuku-Gama system of the Eastern Central Highlands of New Guinea. The dataset consists of two parts: (1) GAMAPOS representing alliance ("rova") relations, and (2) GAMANEG for antagonistic ("hina") relations \cite{read_cultures_1954}.

The college A, B, and C datasets are constructed with respect to the relationships between three different groups of college girls at an Eastern college. Each group consists of approximately 20 members who lived together for at least four months. The girls in each house were asked to provide evaluations about other girls in the same house based on a range of behavioral characteristics. These data are then converted into matrices of choice and rejection for each house \cite{lemann1952group}.

Sampson monastery network represents positive and negative relationships between 18 monks (novices) in a monastery. To construct the social network, each monk was asked to rank his top three choices with respect to four positive social relationships. In addition, they were asked to create the same list for negative relationships. The data is collected at five different times, but the group we analyze were present only in three data collection times \cite{sampson1969novitiate}. 

The Newcomb’s fraternity dataset represent the relationships between 17 boys living in a pseudo-dormitory. The members of the dormitory were asked to rank their relationship with other people over the period of 15 weeks. The top and bottom preference rankings are assumed to represent positive and negative relationships respectively \cite{newcomb_acquaintance_1961}.

The Philosophers network is constructed based on Randal Collin's in his seminal work ``The Sociology of Philosophies'' \cite{collins2009sociology}, in which he studies various types of philosophical thoughts in ancient Greece, China, Japan, India, the medieval Islamic and Jewish world, medieval Christendom, and modern Europe. The book represents division and conflicts among philosophers based on acquaintanceship relationships and relationships between masters and pupils.  

\subsection*{From the micro-level to the macro-level of balance}
The graph-theoretical background of balance theory shows a simple, but often disregarded limitation for evaluating balance merely based on triads of a network: balance of a network's triads is not a sufficient condition for balance of the network. Some researchers have circumvented this issue in undirected signed graphs by arguing that cycles longer than 3 are not particularly important \cite{bonacich_introduction_2012} and continued using triads on the basis that their balance is a necessary condition for balance of a network. In digraphs, however, both of these premises fall apart which accentuates the problem. In an unbalanced digraph, all semicycles of length 3 and above could be balanced if an unbalanced semicycle of length 2 exists. Balance in transitive semicycles of length 3 is not the only condition for balance of a signed digraph. In our proposed methodology, we have also used the line index of balance which provides a complementary perspective on balance compared to what is achievable by using only semicycles and triads. A key connection between our micro-level and macro-level measures is that every unbalanced triad has at least one frustrated edge \cite{aref2016exact}, and that the frustration index equals the minimum number of unbalanced fundamental cycles induced over all spanning trees of the graph \cite{iacono_determining_2010}. 

\subsection*{Theoretical underpinnings of our approach}
Our multilevel framework of structural balance evaluation is motivated by an extensive pool of literature from network science \cite{harary1959measurement,doreian_partitioning_1996,aref2015measuring,aref2017balance} and organizational sciences \cite{crawford2013intra-inter,cohen2008war,lau1998faultline-theory}. The literature recognizes that networks inherently consist of multiple levels \cite{monge2003theories}, in which different social forces and structural processes co-occur that could influence balance dynamics. We begin our multilevel framework with triadic analysis, primarily because triads is often assumed to be building blocks of networks \cite{wasserman1994social} where structural dynamics involving signs and directionality take place. Indeed, earliest formulation of structural balance was made by Heider \cite{heider1946attitudes} at the triadic level (2 persons and a common object). Heider developed a typology of triads which included four sign configurations of balanced triads (with 0 or 2 negative edges), and four sign configurations of unbalanced triads (with 1 or 3 negative edges). Latter efforts expanded these triadic configurations to consider directionality of edges \cite{rapoport1983mathematical,levi1969elementary,sherwin1971introduction,wasserman1994social}. This subset of literature leveraged semicycles and posited that a signed directed triad is ``balanced if and only if all semicycles have positive signs'' \cite{wasserman1994social}. Johnsen \cite{johnsen1986structure}, followed by Holland and Leinhardt \cite{holland1970method}, then developed the triad census that contained all possible configurations of balanced and unbalanced triads, considering both signs and directions of edges in semicycles. 

At the same time, several network science scholars extended structural balance measures that included groupings beyond three people (a triad). Cartwright and Harary \cite{cartwright_structural_1956} derived the \textit{structure theorem} that posited a signed network to be balanced if it is possible to subdivide all nodes into two sets in ways that all positive edges occur within the same set, and all negative edges occur between sets. Doreian and Mrvar \cite{doreian_partitioning_1996} proposed a similar framework to partition nodes into distinct \textit{plus-sets}. Davis and Leinhardt \cite{davis1967} then proposed a ranked clusterability model that grouped dyads together in the same cluster if they both have positive and reciprocated edges, and separated dyads to two clusters if they both have negative and reciprocated edges. The structural notion of positive ties within subgroup and negative ties between subgroups was also widely examined in studies of teams \cite{crawford2013intra-inter,cohen2008war} and organizational communication \cite{lau1998faultline-theory}. A study of team trust by Ashleigh and Stanton \cite{ashleigh2001trust}, for example, found that most teams have stronger trust scores at the intra-team level than the inter-team level, primarily because members have already established a strong in-group identity, and thus view working with the out-group members as a competitive task. Lau and Murighan \cite{lau1998faultline-theory} observed the emergence of ``faultlines'' in work-groups based on demographic attributes such as gender, age, and occupational roles. These faultlines produced homogeneous subgroups where individuals have strong internal cohesion, but weak external connections with other subgroups. 

While triad-level and subgroup-level analyses capture the fine-grained dynamics that may influence balance between groups of nodes, there is a need to examine balance at the network-level to capture structural forces that influence balance beyond the groups and at a holistic level. Harary \cite{harary1959measurement} proposed line index of balance to determine the smallest number of edges whose change of sign would result in a balanced network. The normalized version of this measure suggested in \cite{aref2015measuring} expands the concept of a network's balance from being a binary state (either balanced or unbalanced) to a continuous scale for the level of balance according to which a network can be partially balanced \cite{aref2015measuring,aref2017balance}. 

\subsection*{Notes on common properties of the signed social networks}

As a result of conducting macro-level balance assessment by computing its frustration index, signed networks will fall into one of three possible categories of partial balance: (1) close to balance, (2) random pattern of signed ties, and (3) far away from balance. In the first group, the frustration index is relatively small compared to its theoretical upper bound $m/2$ \cite{aref2017computing}, which leads to $F(G)$ being in the range of $(0.7,1]$. Therefore, the network is just a few edges away from balance. In the second group, the frustration index is neither small nor large when compared to $m/2$, such that $F(G)$ takes values around $0.5$. Random signed graphs were observed to show such behavior \cite{aref2015measuring}, where the frustrated index is roughly half of its maximum possible value $m/2$. In the third group, $F(G)$ takes values from $[0,0.3)$, because signed ties are arranged such that the overall network is far away from balance, and a large number of edges (compared to $m/2$) would need to be removed to achieve balance. An extreme example is a complete signed digraph in which every pair of nodes is connected by one positive and one negative tie. In such a network, at least half of the edges would need to be removed to achieve a balance network ($L(G)=m/2, F(G)=0$).

Given that a large majority of the networks we analyzed represent high values of balance in different levels, one may wonder if the networks were selected only if they show consistency with balance theory. This is not the case as we have included all publicly available signed directed network data that we have found from social domain. Therefore, observing high partial balance in all these networks despite their difference in context is yet another empirical confirmation for balance theory \cite{facchetti_computing_2011,aref2017balance} and specifically how signed social networks represent a high level of partial (as opposed to total) balance \cite{aref2015measuring}.

Another possible concern could be that our methods of measurement are not calibrated and therefore only provide values in the higher half of the range $[0,1]$ regardless of what the network is. This is not the case according to the axiomatic comparison of balance measures in \cite{aref2015measuring} and our testing of these measures on a different static network. \textit{Chess network} \cite{kunegis2013konect} is a signed digraph produced from real data which represent chess players as nodes and the game outcomes as signed directed edges from the white player to the black player with a positive (negative) sign for white victory (defeat). While this network comes form a real data-generation process, there is no social meaning to the ties and we would not expect the network to represent high partial balance (consistency with balance theory). Confirming our intuition, triad level balance is measured at $0.47$ and the network level balance is $0.4$ for the Chess network (both values showing a random pattern in signed ties) which show that the two measures could distinguish networks by providing high or low values as required.

\subsection*{Difference and relevance of micro- and macro-level balance}
Despite the support for the argument that balance in micro and macro levels are not generally the same property measured at different levels, there are special cases where the micro and macro balance cannot be separated at all as if they are indeed the same property. A fully connected signed network (which has a density and a clustering coefficient of 1) is balanced if and only if its triads are balanced \cite{zaslavsky_mathematical_2012}. Therefore, one could argue that in complete networks, not only the aggregation of balance in micro-level represents the balance of the overall structure, but balance at either micro- or macro-level is not mathematically possible without balance at the other level.

\subsection*{Uniqueness and multiplicity of optimal partitions}
An essential part of our analysis of the networks has been the optimal partitions which is the first step for the meso-level and macro-level evaluation of balance. Previous studies on frustration index \cite{aref2017computing,aref2016exact,aref2017balance,aref2020detecting} do not discuss the multiplicity of optimal partitions. We configure the solving strategy of the mathematical model in Eq.\ \eqref{eq1} to not just find one optimal solution, but systematically search through the space of solutions (feasible space) and find all optimal solutions. 
This involves finding all all partitions whose frustration count equals the frustration index.

\begin{figure}[ht] 
	\centering
	\minipage{0.5\textwidth}
	\includegraphics[width=\linewidth]{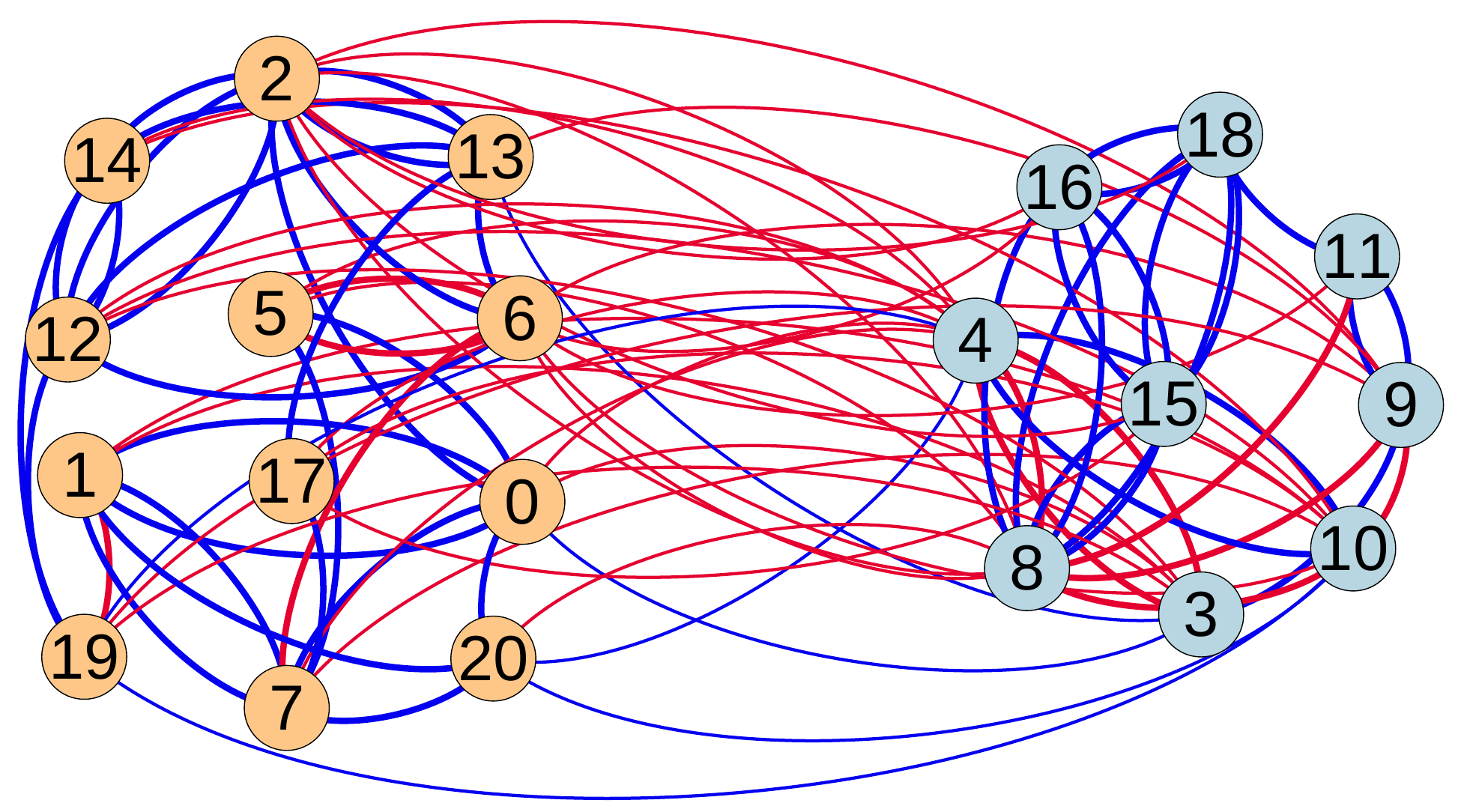}
	\caption*{Optimal Partition $P_1^*$}
	\endminipage\hfill
	\minipage{0.5\textwidth}
	\includegraphics[width=\linewidth]{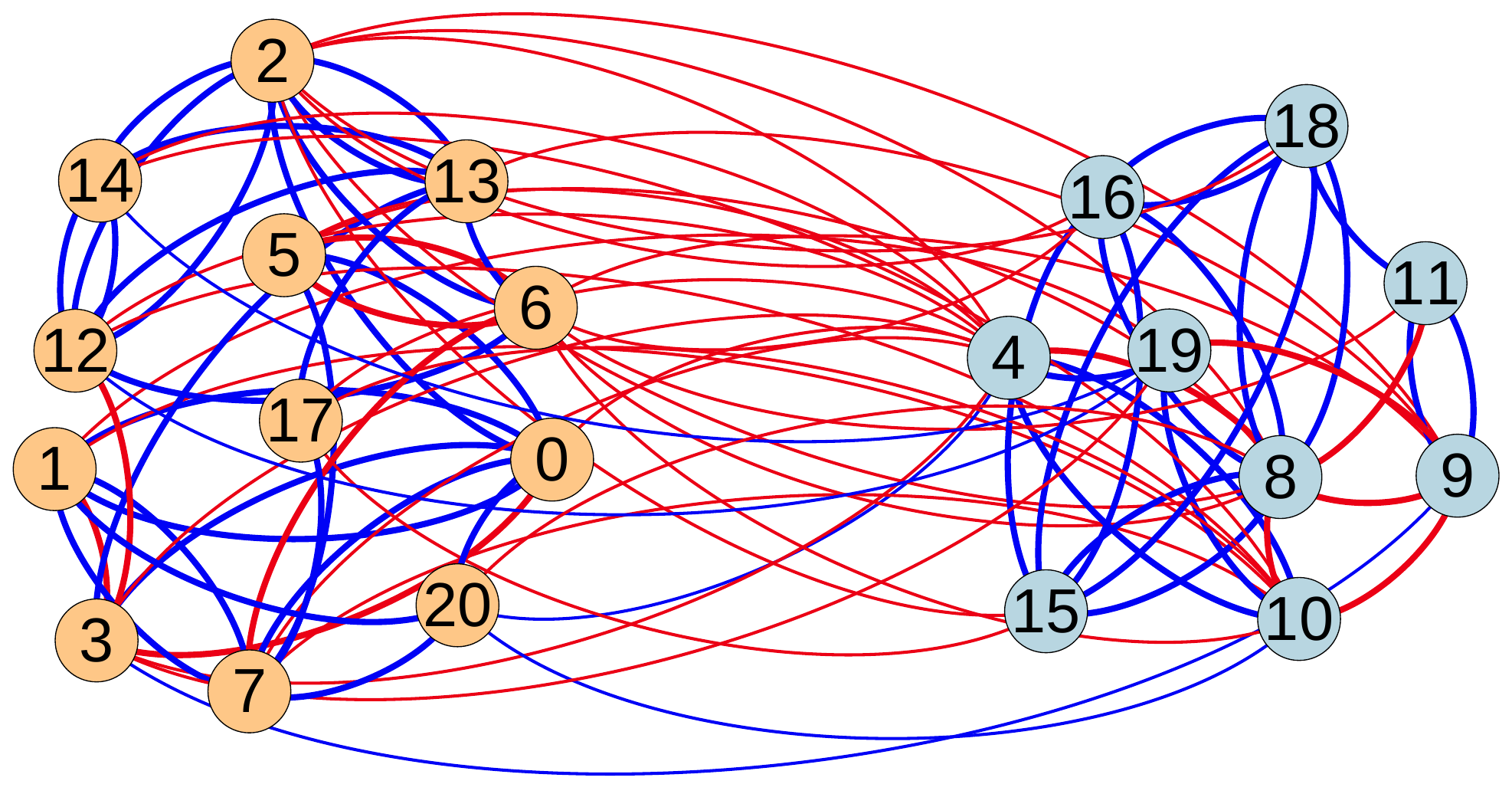}
	\caption*{Optimal Partition $P_2^*$}
	\endminipage\hfill
	\minipage{0.5\textwidth}%
	\includegraphics[width=\linewidth]{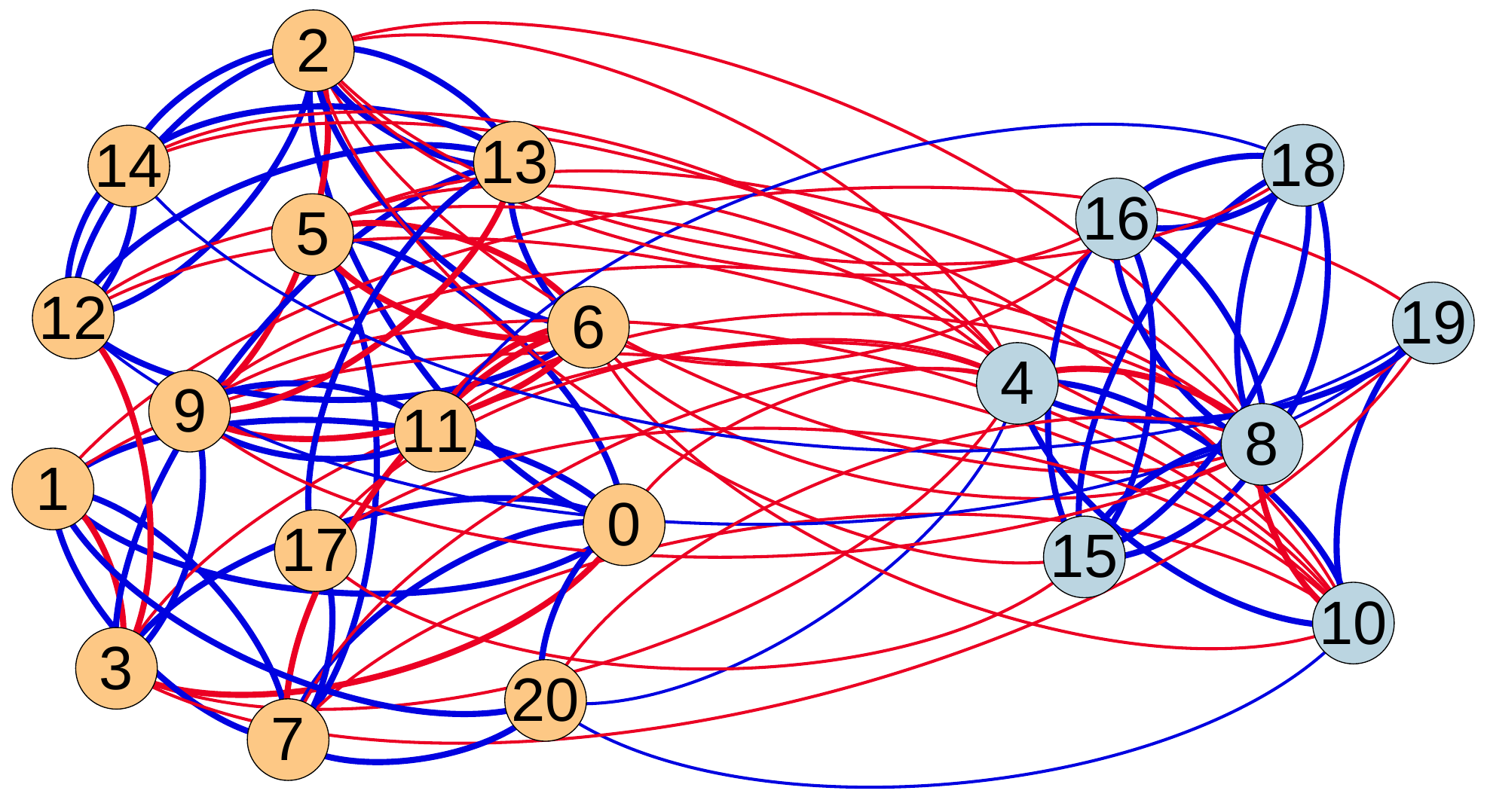}
	\caption*{Optimal Partition $P_3^*$}
	\endminipage
	\caption{Three distinct optimal partitions achievable for college preferences in House A. Direction of arcs are clockwise. Blue arcs are positive and red arcs are negative.}
	\label{fig:FigS1}
\end{figure}

This further analysis requires intensive computations and can only be done for relatively small network. Solving our proposed optimization model with this new search strategy allows us to confirm that the optimal solution is unique in \textit{highland tribes}, all time-frames of \textit{Sampson monastery} and \textit{college preferences in houses B and C}. In contrast, we find that for \textit{college preference house A}, optimal solution is not unique and has a multiplicity of 3 (there are three partitions whose frustration count equals the frustration index). In Figure \ref{fig:FigS1}, the three optimal partitions of \textit{college preference house A} are visualized.

While the number of frustrated edges in all these three optimal partitions are equal, there could be differences on the composition of positive and negative edges among the frustrated edges in optimal partitions which in turn could impact our measurements of cohesiveness and divisiveness.

\begin{table}[ht]
	\caption{Multiple optimal partitions for college preferences in House A and values of cohesiveness and divisiveness}
	\label{tab:TabS1}
	\centering
	\begin{tabular}{llll} \hline
		Optimal partition & $X^*$ & Cohesiveness $C(P_i^*)$ & Divisiveness $D(P_i^*)$ \\ \hline
		$P_1^*$ & \{4,8,10,15,16,18,\ 3,9,11\} & 0.804 & 0.842 \\
		$P_2^*$ & \{4,8,10,15,16,18,\ 9,11,19\}& 0.793& 0.861 \\
		$P_3^*$ & \{4,8,10,15,16,18,\ 19\}     & 0.793 &  0.861\\ \hline
	\end{tabular}
\end{table}

\FloatBarrier

In \textit{college preference house A} network, the subset assignment of all nodes are consistent across multiple partitions except for four nodes: 3,9,11,19. As it can be seen in Table \ref{tab:TabS1} and Figure \ref{fig:FigS1} that these four nodes belong to different subsets across the three optimal partitions while nodes 9 and 11 are always together and nodes 3 and 19 are always apart.

The values of cohesiveness and divisiveness are the same for $P_2^*,P_3^*$ because their composition of positive and negative frustrated edges are the same, but for $P_1^*$ there is a small difference in the two measurements. The three optimal partitions of \textit{college preference house A}, leads to a small standard deviation of $0.006$ between the cohesiveness values and a standard deviation of $0.011$ between the divisiveness values.

This observation shows that there are networks in which the position of some nodes could be rather precarious across the two optimal subsets. For nodes 9 and 11 (connected by two positive arcs), membership to either of the two subsets is possible as long as they stick together. For nodes 3 and 19 (connected by a negative arc), optimal group assignments place them in different groups, but they all keep the two nodes separated. No possible benefit (in terms of reduction in frustration count) exists in the network that would offset the cost of separating 9 and 11 or the cost of placing 3 and 19 together. 




\FloatBarrier

\section*{Detailed Materials and Methods}
\subsection*{Processing of secondary data into network data}
In this subsection, we provide detailed descriptions of data preprocessing for balance analysis of static networks, temporal networks, and multilayer network. All datasets are publicly available, and most are available through major network data repositories such as UCINET IV, Stanford Network Analysis Project (SNAP), and Index of Complex Networks (ICON). The exceptions were \textit{Sampson's affect data} \cite{sampson1969novitiate}, \textit{College Houses A,B,C} \cite{lemann1952group}, and \textit{Philosophers network} which were not in digitized form originally. We digitized Sampson's affect data from Tables in \cite[p. 160-161]{doreian_partitioning_1996} considering all the edges. We reconstructed the \textit{College Houses} networks from sociomatrix Tables in \cite[p. 18-20]{lemann1952group} such that all choices and rejections that had a weight of $-4,-3,+3,+4$ were considered as directed signed edges. Collins \textit{Philosophers} network was digitized by Wouter de Nooy and provided to the authors.  

\textbf{Static networks:} For four networks (\textit{Reddit}, \textit{Wikipedia election}, \textit{Bitcoin OTC}, \textit{Bitcoin Alpha}) we kept the original datasets without much preprocessing, given that the edge-lists were directed, and contained signed attributes ($-1$ or $+1$ in \textit{Reddit} and \textit{Wikipedia}, and $-10$ to $+10$, excluding $0$, in \textit{Bitcoin OTC} and \textit{Bitcoin Alpha}). The strength of the edges are not considered and only the signs were imported for creating the networks. For \textit{College Houses} \textit{A, B, C}, the sign attribute contained values from $-4$ to $+4$. We only include edges that contain scores of $-3$ and $-4$, which we relabel to $-1$. and scores of $+3$ and $+4$, which we relabeled to $+1$. Lastly, for the \textit{Highland tribes} network, we combined two edge-lists, \textit{GAMAPOS} ($+1$) and \textit{GAMANEG} ($-1$), which contains positive and negative relations among tribes, respectively. We then symmetrized the edge-list to convert the network from an undirected to a directed representation. 

\textbf{Temporal networks:} 
For \textit{Sampson's affect}, we use the the data we had digitized from Tables in \cite[p. 160-161]{doreian_partitioning_1996} which has weight range from $-1$ to $-3$ and from $+1$ to $+3$, but we discard the strength of the edges and load the edges only based on their signs.

For \textit{Newcomb's Fraternity}, the original dataset contains 15 matrices on the preference ranking from 17 men. Each person provides a score from 1-16 for every other person in the fraternity house, in decreasing order of their preference for friendship with that person. To create a directed signed network, we filter the data to only include the reported top 3 ranks and bottom 3 ranks (which we consider as positive and negative edges respectively). We do not consider different strengths for the edges. For \textit{Newcomb's fraternity} data, we create signed graphs with edges based on the three least (most) preferred nominations of each person as their enemies (friends). For negative ties, in particular, this choice may be contestable even though it is consistent with the balance analysis of the same data in \cite{doreian1996brief}. Our preliminary post-hoc analysis that uses different choices for inferring signed ties (e.g., top and bottom 1 ranks, top and bottom 2 ranks) reveals notably different balance values. Therefore, further investigations on modeling signed network data are required to better understand and interpret signed networks inferred from preference ranking data.

\textbf{Multilayer network:} For \textit{Philosophers network}, we considered edges with weights of $-2$ and $+2$, representing negative and positive edges between philosophers, respectively. We discarded edges with weight of $1$, as they represent ``probable ties'' according to \cite{collins2009sociology}. We then symmetrized all edges of the network in the two layers (master-pupil and acquaintanceship), and combined the two layers to create a flattened representation of the multilayer network. 

After all network edge-lists were constructed, we pre-processed the data using \textit{pandas} in Python. This step included calculation of triad-level balance and descriptive metrics such as global clustering coefficient, density, and triad census.  

\subsection*{Obtaining optimal partition of directed signed graphs}

Following recently developed computational methods \cite{aref2016exact}, we formalize the process for obtaining a globally optimal partition for a signed digraph as a binary linear programming model in Eq.\ \eqref{eq1} which computes the frustration index of input digraph $G$ in its optimal objective function.
\begin{equation}
\label{eq1}
\begin{split}
\min_{x_i, f_{ij}} Z &= \sum\limits_{(i,j) \in E}  f_{ij}  \\
\text{s.t.} \quad
f_{ij} &\geq x_{i}-x_{j} \quad \forall (i,j) \in E^+ \\
f_{ij} &\geq x_{j}-x_{i} \quad \forall (i,j) \in E^+ \\
f_{ij}  &\ge  x_{i} + x_{j} -1 \quad \forall (i,j) \in E^- \\
f_{ij}  &\ge  1-x_{i} - x_{j}  \quad \forall (i,j) \in E^- \\
x_{i} &\in \{0,1\} \quad  \forall i \in V \\
f_{ij} &\in \{0,1\} \quad \forall (i,j) \in E 
\end{split}
\end{equation}

The optimization model formulated in Eq.\ \eqref{eq1} has one binary variable for each vertex and one binary variable for each edge of the digraph $G$. Binary variable $x_{i}$ is associated with node $i$ and represents the subset that node $i$ belongs to (whether $i \in X$ or not as a binary value). Binary variable $f_{ij}$ is associated with directed edge $(i,j)$ and represents its frustration state (whether directed edge $(i,j)$ is frustrated or not as a binary value).

The objective function is the minimum of the total count of frustrated edges over the set of edges $E$. The first two constraints ensure that positive edges $(i,j) \in E^+$ are counted as frustrated if their endpoints $i$ and $j$ belong to different subsets. Similarly, the next two constraints handle the frustration of negative edges $(i,j) \in E^-$ when their endpoints $i$ and $j$ belong to the same subset. 
We solve the optimization model in Eq.\ \eqref{eq1} using \textit{Gurobi} solver \cite{gurobi} (version 9.0) in \textit{Python} and for large networks we follow the two-step method discussed in \cite{aref2020detecting}.

\subsection*{Computational analysis and solving the optimization model}

The code for all the computational analysis and the optimization model will be made available on a Github repository at \url{https://github.com/saref/mutlilevel-balance} 
once this paper is published.

The code for multilevel evaluation of balance including the optimization models are distributed under an Attribution-NonCommercial-ShareAlike 4.0 International (CC BY-NC-SA 4.0) license. This means that one can use these algorithms for non-commercial purposes provided that they provide proper attribution for them by citing \cite{aref2017computing,aref2016exact} and the current article. Copies or adaptations of the algorithms should be released under the similar license.

For the computational analysis at the micro level, our algorithm lists graph triads and checks their transitivity. If all semicycles of a triad are transitive, then their balance is evaluated and provided as the output.

The proposed optimization model can be solved by any mathematical programming solver which supports 0/1 linear programming (binary linear) models. In the Github repository \url{https://github.com/saref/multilevel-balance}, we explain using Gurobi solver (version 9.0) for solving the proposed model. 

Our proposed optimization algorithm is developed in Python 3.7 based on the mathematical programming models developed in \cite{aref2017computing,aref2016exact,aref2020detecting}, but it computes the frustration index of a directed signed graph.

The following steps outline the process for academics to install the required software (\textit{Gurobi} solver \cite{gurobi}) on their computer to be able to run the optimization algorithms:

\begin{enumerate}
	\item 
	Download and install Anaconda (Python 3.7 version) which allows you to run a Jupyter code. It can be downloaded from \url{https://www.anaconda.com/distribution/}. Note that you must select your operating system first and download the corresponding installer.
	
	\item 
	Register for an account on \url{gurobi.com/registration-general-reg/} to get a free academic license for using Gurobi. Note that Gurobi is a commercial software, but it can be registered with a free academic license if the user is affiliated with a recognized degree-granting academic institution. This involves creating an account on Gurobi website to be able to request a free academic license in step 5.
	
	\item 
	Download and install Gurobi Optimizer (versions 9.0 and above are recommended) which can be downloaded from \url{https://www.gurobi.com/downloads/gurobi-optimizer-eula/} after reading and agreeing to Gurobi's End User License Agreement.
	
	\item
	Install Gurobi into Anaconda. You do this by first adding the Gurobi channel to your Anaconda channels and then installing the Gurobi package from this channel.
	
	From a terminal window issue the following command to add the Gurobi channel to your default search list
	
	\begin{verbatim}
	conda config --add channels 
	http://conda.anaconda.org/gurobi
	\end{verbatim} 
	
	Now issue the following command to install the Gurobi package
	
	\begin{verbatim}
	conda install gurobi
	\end{verbatim}
	
	\item 
	Request an academic license from \url{gurobi.com/downloads/end-user-license-agreement-academic/} and install the license on your computer following the instructions given on Gurobi license page.
	
	Completing these steps is explained in the following links (for version 9.0):
	
	for Windows \url{https://www.gurobi.com/documentation/9.0/quickstart_windows/installing_the_anaconda_py.html},
	
	for Linux \url{gurobi.com/documentation/9.0/quickstart_linux/installing_the_anaconda_py.html}, and
	
	for Mac \url{gurobi.com/documentation/9.0/quickstart_mac/installing_the_anaconda_py.html}.
	
	After following the instructions above, open Jupyter Notebook which takes you to an environment (a new tab on your browser pops up on your screen) where you can open the main code (which is a file with .ipynb extension).
	
\end{enumerate}

\subsection*{Additional numerical results for the networks}
Table \ref{tab:TabS2} provide all the numerical results and additional measurements produced for the network.

\FloatBarrier

\begin{sidewaystable}\centering
	\caption{Additional numerical results for the networks}
	\begin{tabular}{llllllllllllll}
		\hline
		Dataset            & $n$   & $m$    & $m^+$  & $m^-$ & B.T.   & U.T.   & T(G)  & C.C.     & Density  & $L(G)$ & $F(G)$ & $C(P^*)$ & $D(P^*)$ \\
		\hline
		Tribes             & 16    & 116    & 58     & 58    & 59     & 9      & 0.87  & 0.527    & 0.483    & 14     & 0.759  & 0.806    & 1        \\
		Reddit             & 18313 & 120792 & 111891 & 8901  & 237148 & 99546  & 0.704 & 6.30E-02 & 3.60E-04 & 8511   & 0.859  & 0.936    & 0.096    \\
		Wikipedia          & 7118  & 103675 & 81318  & 22357 & 426215 & 141380 & 0.751 & 5.30E-02 & 2.00E-03 & 15045  & 0.71   & 0.869    & 0.765    \\
		Bitcoin Alpha      & 3783  & 24186  & 22650  & 1536  & 11649  & 2141   & 0.845 & 6.40E-02 & 1.70E-03 & 1098   & 0.909  & 0.96     & 0.781    \\
		Bitcoin OTC        & 5881  & 35592  & 32029  & 3563  & 19447  & 2969   & 0.866 & 4.50E-02 & 1.00E-03 & 1644   & 0.908  & 0.96     & 0.871    \\
		House A            & 21    & 94     & 51     & 43    & 46     & 11     & 0.807 & 0.392    & 0.224    & 17     & 0.638  & 0.793    & 0.861    \\
		House B            & 17    & 83     & 41     & 42    & 24     & 22     & 0.522 & 0.398    & 0.305    & 19     & 0.542  & 0.739    & 0.811    \\
		House C            & 20    & 81     & 41     & 40    & 26     & 3      & 0.896 & 0.271    & 0.213    & 5      & 0.877  & 0.909    & 0.973    \\
		Sampson T2         & 18    & 104    & 55     & 49    & 41     & 19     & 0.683 & 0.349    & 0.34     & 21     & 0.596  & 0.827    & 0.769    \\
		Sampson T3         & 18    & 105    & 57     & 48    & 52     & 24     & 0.684 & 0.445    & 0.343    & 20     & 0.619  & 0.825    & 0.792    \\
		Sampson T4         & 18    & 103    & 56     & 47    & 49     & 16     & 0.754 & 0.412    & 0.337    & 14     & 0.728  & 0.85     & 0.884    \\
		Fraternity 00      & 17    & 102    & 51     & 51    & 38     & 37     & 0.507 & 0.376    & 0.375    & 26     & 0.49   & 0.745    & 0.745    \\
		Fraternity 01      & 17    & 102    & 51     & 51    & 35     & 23     & 0.603 & 0.406    & 0.375    & 22     & 0.569  & 0.746    & 0.837    \\
		Fraternity 02      & 17    & 102    & 51     & 51    & 26     & 28     & 0.481 & 0.406    & 0.375    & 25     & 0.51   & 0.741    & 0.771    \\
		Fraternity 03      & 17    & 102    & 51     & 51    & 32     & 18     & 0.64  & 0.376    & 0.375    & 25     & 0.51   & 0.741    & 0.771    \\
		Fraternity 04      & 17    & 102    & 51     & 51    & 47     & 36     & 0.566 & 0.435    & 0.375    & 27     & 0.471  & 0.731    & 0.74     \\
		Fraternity 05      & 17    & 102    & 51     & 51    & 44     & 41     & 0.518 & 0.492    & 0.375    & 24     & 0.529  & 0.745    & 0.787    \\
		Fraternity 06      & 17    & 102    & 51     & 51    & 50     & 53     & 0.485 & 0.484    & 0.375    & 25     & 0.51   & 0.76     & 0.75     \\
		Fraternity 07      & 17    & 102    & 51     & 51    & 50     & 49     & 0.505 & 0.525    & 0.375    & 25     & 0.51   & 0.724    & 0.795    \\
		Fraternity 08      & 17    & 102    & 51     & 51    & 40     & 53     & 0.43  & 0.543    & 0.375    & 26     & 0.49   & 0.727    & 0.766    \\
		Fraternity 10      & 17    & 102    & 51     & 51    & 40     & 37     & 0.519 & 0.461    & 0.375    & 26     & 0.49   & 0.736    & 0.755    \\
		Fraternity 11      & 17    & 102    & 51     & 51    & 37     & 36     & 0.507 & 0.467    & 0.375    & 22     & 0.569  & 0.796    & 0.774    \\
		Fraternity 12      & 17    & 102    & 51     & 51    & 38     & 34     & 0.528 & 0.486    & 0.375    & 22     & 0.569  & 0.784    & 0.784    \\
		Fraternity 13      & 17    & 102    & 51     & 51    & 44     & 46     & 0.489 & 0.516    & 0.375    & 21     & 0.588  & 0.788    & 0.8      \\
		Fraternity 14      & 17    & 102    & 51     & 51    & 44     & 38     & 0.537 & 0.475    & 0.375    & 21     & 0.588  & 0.8      & 0.788    \\
		Fraternity 15      & 17    & 102    & 51     & 51    & 50     & 37     & 0.575 & 0.498    & 0.375    & 23     & 0.549  & 0.769    & 0.78     \\
		Master-pupil layer & 712   & 1354   & 1264   & 90    & 18     & 1      & 0.947 & 0.038    & 0.003    & 4      & 0.994  & 0.998    & 0.988    \\
		Acquaintance layer & 346   & 660    & 474    & 186   & 72     & 2      & 0.917 & 0.091    & 0.006    & 6      & 0.982  & 0.996    & 0.979    \\
		Philosophers flat  & 855   & 2010   & 1736   & 274   & 78     & 19     & 0.804 & 0.089    & 0.003    & 60     & 0.94   & 0.976    & 0.931 \\  \hline
	\end{tabular}
	\\
	B.T.: Balanced triads, U.T.: Unbalanced triads, C.C.: Clustering coefficient
	\label{tab:TabS2}
\end{sidewaystable}

\FloatBarrier

\begin{figure}[ht]
	\centering
	\includegraphics[width=\textwidth]{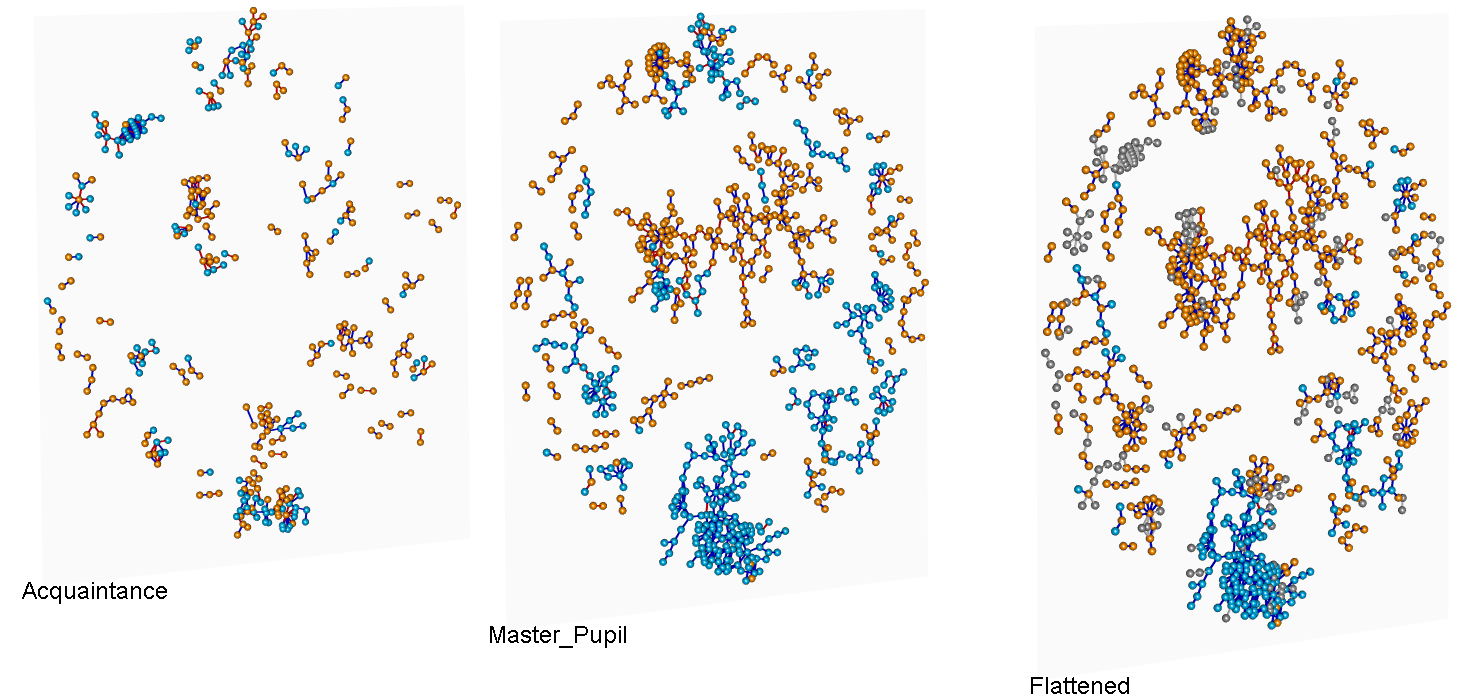}
	\caption{Two layers of the philosophers network and its flattened view}
	\label{fig:FigS2}
\end{figure}

\begin{figure}
	\centering
	\includegraphics[width=\textwidth]{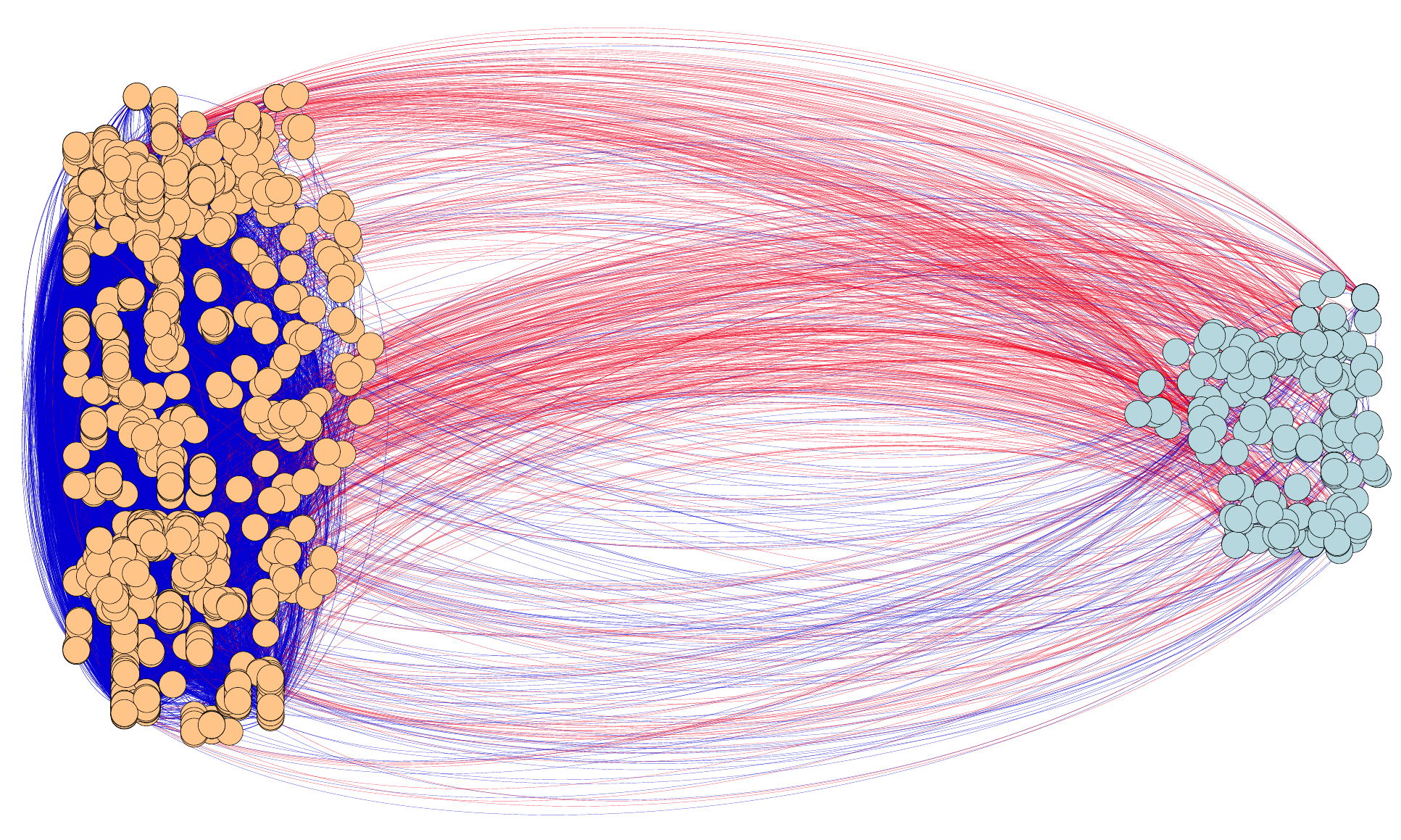}
	\caption{Visualization of Bitcoin-Alpha network and its optimal partition (high cohesiveness, high divisiveness)}
	\label{fig:FigS3}
\end{figure}

\begin{figure}
	\centering
	\includegraphics[width=\textwidth]{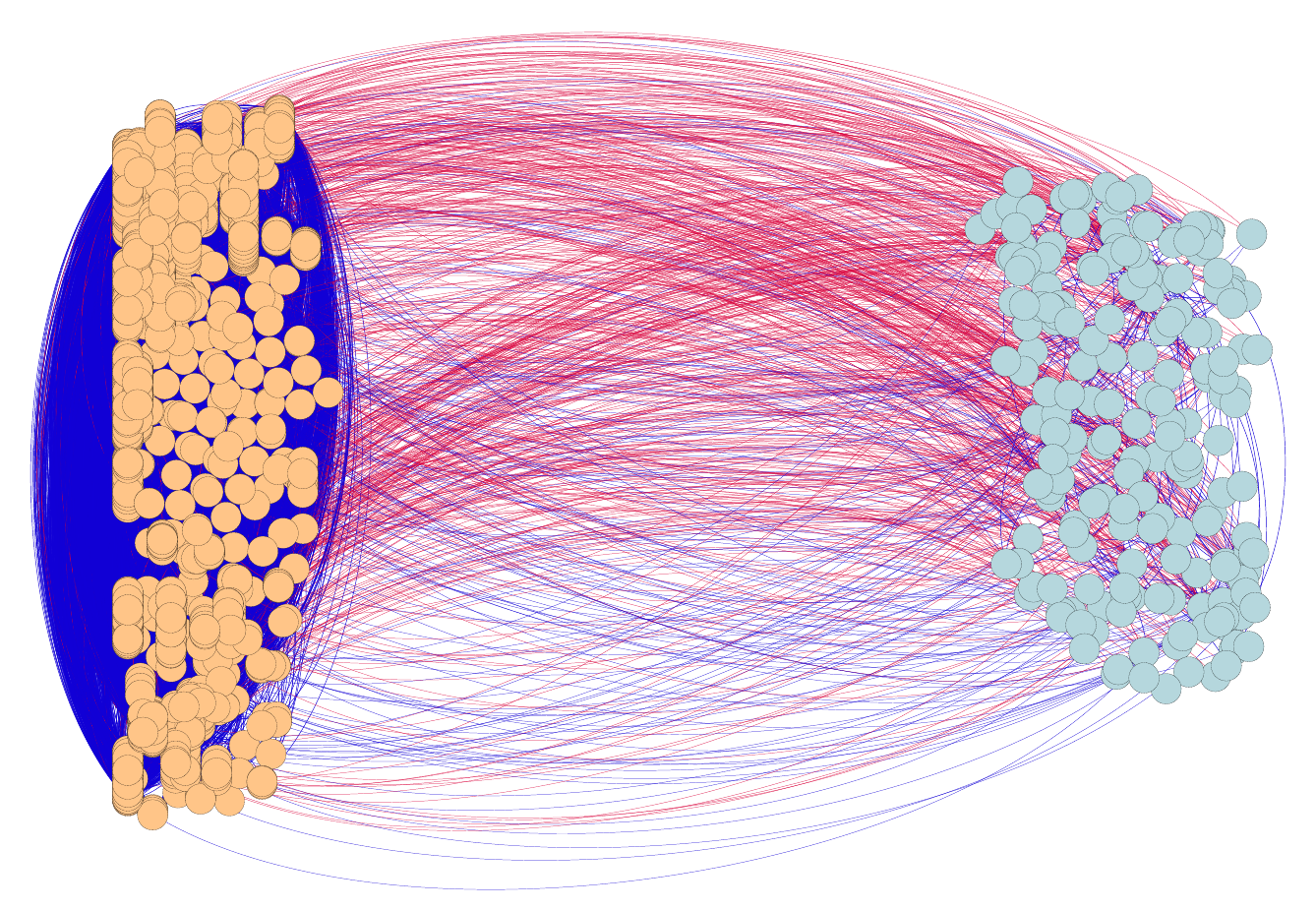}
	\caption{Visualization of Bitcoin-OTC network and its optimal partition (high cohesiveness, high divisiveness)}
	\label{fig:FigS4}
\end{figure}

\begin{figure}
	\centering
	\includegraphics[width=\textwidth]{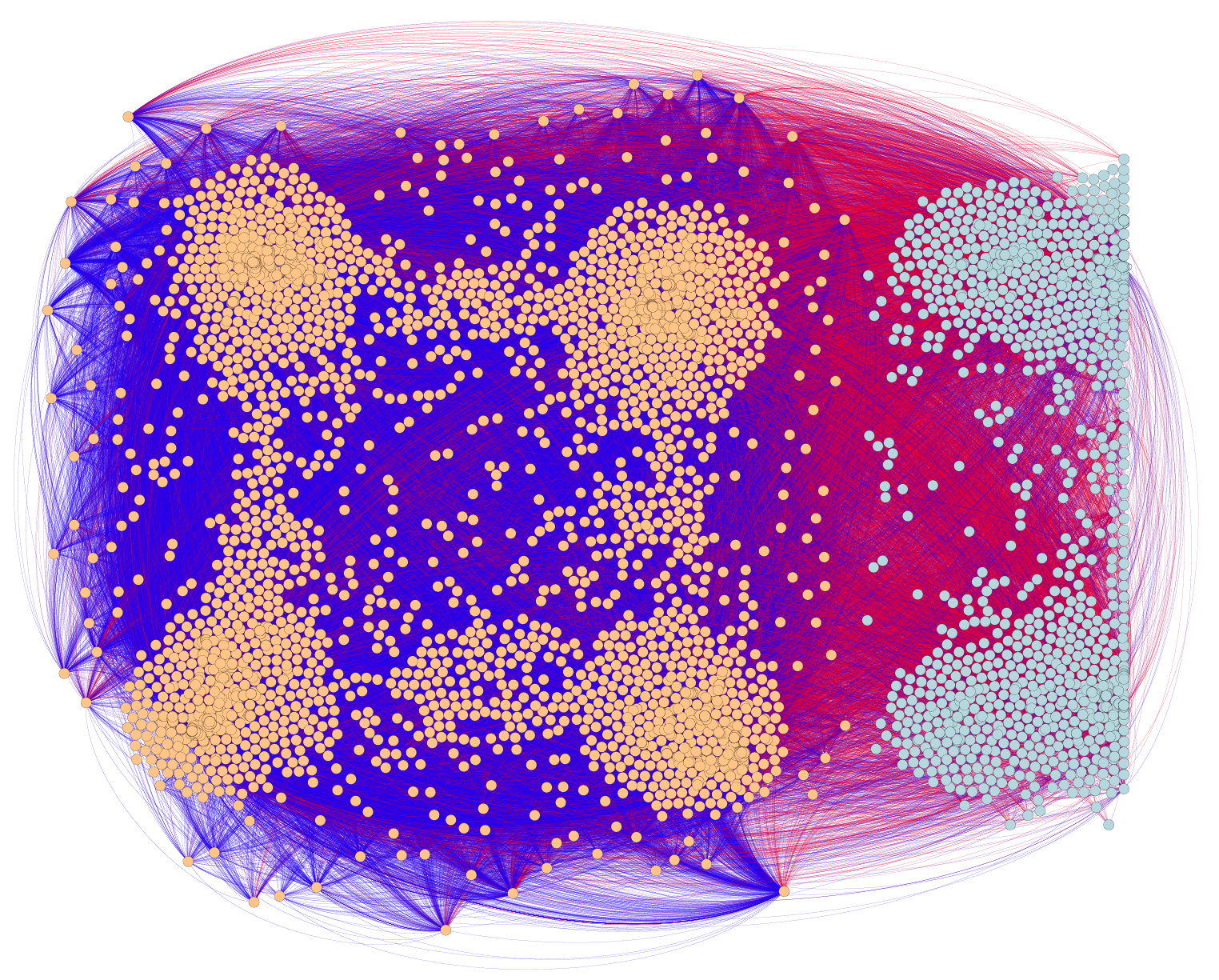}
	\caption{Visualization of Wikipedia network and its optimal partition (high cohesiveness, high divisiveness)}
	\label{fig:FigS5}
\end{figure}

\begin{figure}
	\centering
	\includegraphics[width=\textwidth]{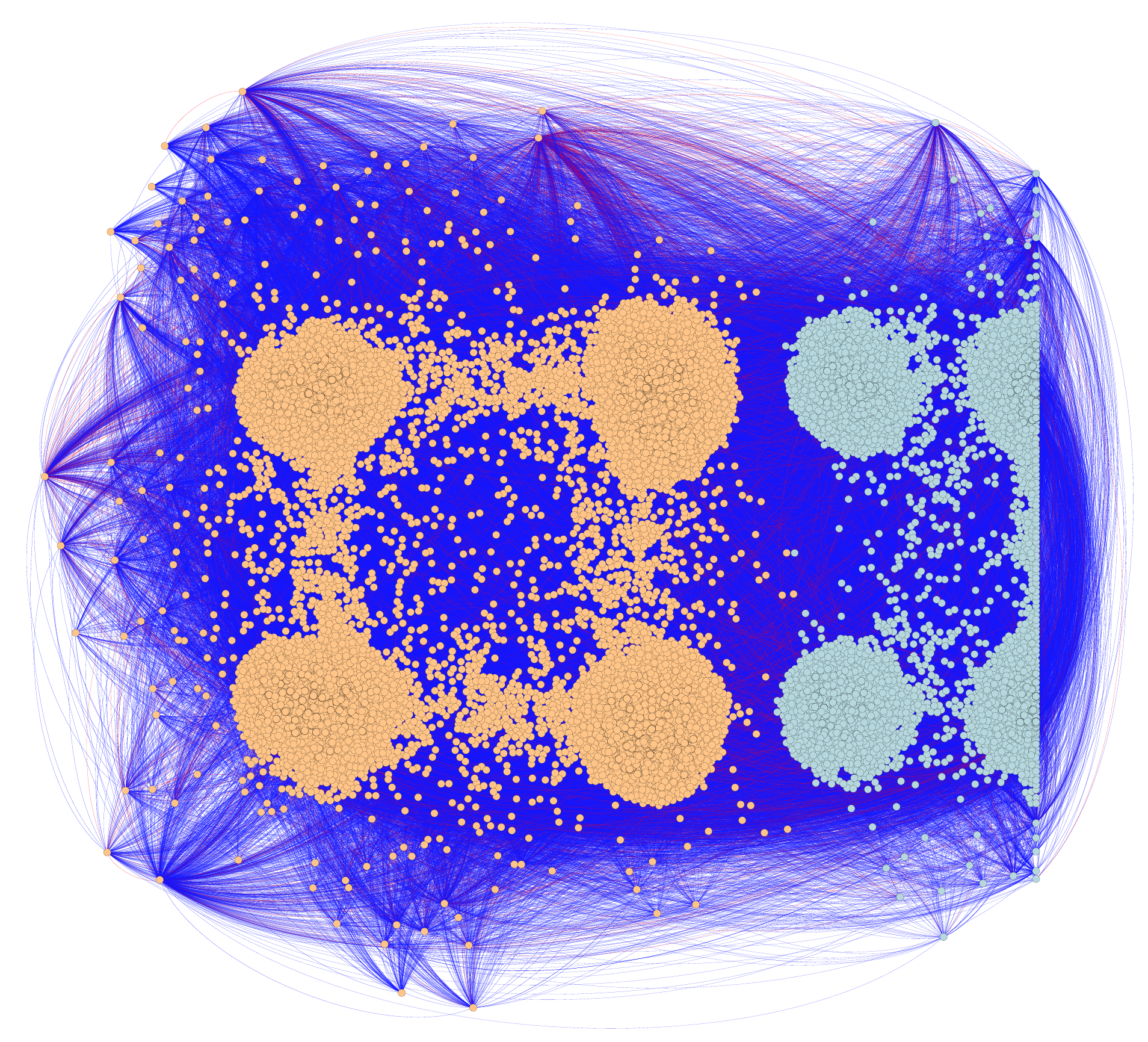}
	\caption{Visualization of Reddit network and its optimal partition (high cohesiveness, low divisiveness)}
	\label{fig:FigS6}
\end{figure}

\subsection*{Visualization of optimal partitions in large networks (Figures \ref{fig:FigS2} to \ref{fig:FigS6})}

Fig.\ \ref{fig:FigS2} shows the \textit{Philosophers} network as a multilayer network which is visualized using \textit{MuxViz} \cite{de2015muxviz}. Figs. \ref{fig:FigS3}--\ref{fig:FigS6} show the optimal partitions in Bitcoin-Alpha, Bitcoin-OTC, Wikipedia, and Reddit respectively. The level of balance at the macro level is high in all these networks. While high macro-level balance is associated with high cohesiveness and high divisiveness (as in Figures \ref{fig:FigS3}--\ref{fig:FigS5}), this does not necessarily have to be the case (as shown in Figure \ref{fig:FigS6}. It can be visually observed that there is very low divisiveness between the subgroups of Reddit in Figure \ref{fig:FigS6}.

\FloatBarrier

\subsection*{Movie: animated versions of optimal partitions in temporal and multilayer networks}

Animated versions of the optimal partitions in temporal networks are available online at \\ \url{https://saref.github.io/SI/ADRD2020/Sampson_partitions.mov}
for Sampson affect data and at \\ \url{https://saref.github.io/SI/ADRD2020/Newcomb_partitions.mov}
for Newcomb fraternity network. The animated visualizations show how the subgroup membership of some of the nodes change over time due to the changes in the edges which in turn alter the optimal partition. The colors used for the nodes represent their optimal group assignemnt in the first time-frame of the temporal network.

Animated version of the optimal partitions in the multilayer network of philosophers is available online at \\ \url{https://saref.github.io/SI/ADRD2020/Collins_partitions.mp4}
The animation shows the two layers of the multilayer network and the flattened network from different orientations.

\subsection*{Dataset: directed signed networks from social domain}
All networks used in this study are available as edge lists in Comma Separated Value format (.csv files) and Geographical Markup Language format (.gml files) accessible in a public \textit{FigShare} data repository \cite{figshare2020directed-signed-graphs}. These data are distributed under a CC-BY 4.0 license. This means that one can use these data provided that they provide proper attribution for them by citing this article and the respective source of original data \cite{lemann1952group,read_cultures_1954,newcomb_acquaintance_1961,sampson1969novitiate,collins2009sociology,west2014exploiting,kumar2016edge,kumar2018community}.

Users of the data in .csv format may notice that for some of the networks, there are values of strength for the edges (values other than $+1, -1$). We have not removed them to keep the dataset more comprehensive. However, in our analysis (and the code which is made publicly available), we load the networks by only considering the signs for the edges and discard any values of strength. Users of the data in .gml format would only see edges with signs of $+1, -1$.

\subsection*{Dataset: network-measurements-and-results.csv}
All numerical results and additional measurements of the networks used in this study are provided in a .csv file at \\ \url{https://saref.github.io/SI/ADRD2020/network-measurements-and-results.csv}. The measurements are described in the column headers which include number of nodes $n$ (order), number of edges $m$ (size),	number of positive and negative edges $m^+,m^-$, number of balanced and unbalanced triads,	fraction of balanced triads $T(G)$, global clustering coefficient, density, triad census, line index of balance $L(G)$, normalized line index  $F(G)$, cohesiveness $C(P^*)$, divisiveness $D(P^*)$, balanced transitive triads by type, and unbalanced transitive triads by type. Each row provides the results and measurements pertaining to a specific network (or a specific time-frame or layer for temporal and multilayer networks) determined by the first column. 

\subsection*{Dataset: optimal-partitions.csv}
The results on optimal partitions (optimal solutions to the optimization model) for all networks are available in a .csv file at \\ \url{https://saref.github.io/SI/ADRD2020/optimal-partitions.csv}.
Each column Refers to a network specified in the first row. Cells in the spreadsheet contain a variable name, index, and its value. For instance, the cell immediately below the header, ``Sampson Affect data at T4'', contains ``x0 : 1'' which means that in time-frame T4 of Sampson network, the $x$ variable associated with node 0, takes value 1 in the optimal solution. Up to four types of variables are reported for each network: $x_i, f_{ij}, s_{ij}, t_{ij}$. Variables $x_i, f_{ij}$ are the decision variables of the optimization model which are associated with the nodes and the edges respectively according to their indices. Variable $s_{ij}$ represent the sign of edge $(i,j)$. Variable $t_{ij}$ represents the combination of sign and optimal situation (internal vs.\ external) for the edge $(i,j)$. There are four combinations for sign and situation denoted by four possible values for $t_{ij}$: positive internal (with value 3), negative internal (with value 1), positive external (with value -1), and negative external (with value -3).

\end{document}